\documentclass[manuscript]{aastex}

\newcommand{\msun}{$\rm M_\odot~$}

\newcommand\nuk[2]{$\rm ^{\rm #2} #1$}

\slugcomment{Accepted for publication in The Astrophysical Journal}

\shortauthors{Chieffi,Limongi} \shorttitle{The evolution of rotating massive stars}

\begin{document}


\title{Presupernova evolution of rotating solar metallicity stars in the mass range 13-120 \msun and their explosive yields.}

\author{Alessandro Chieffi\altaffilmark{1,3} and Marco Limongi\altaffilmark{2,3,4}}

\altaffiltext{1}{Istituto Nazionale di Astrofisica - Istituto di Astrofisica e Planetologia Spaziali, Via Fosso del Cavaliere 100, I-00133, Roma, Italy;
alessandro.chieffi@inaf.it}

\altaffiltext{2}{Istituto Nazionale di Astrofisica - Osservatorio Astronomico di Roma, Via Frascati 33, I-00040, Monteporzio Catone, Italy;
marco.limongi@oa-roma.inaf.it}

\altaffiltext{3}{Centre for Stellar \& Planetary Astrophysics, 
School of Mathematical Sciences, P.O. Box, 28M, Monash University, Victoria 3800, Australia}

\altaffiltext{4}{Kavli Institute for the Physics and Mathematics of the Universe, Todai Institutes for Advanced Study, the University of Tokyo, Kashiwa, Japan 277-8583 (Kavli IPMU, WPI)}

\begin{abstract} 
We present the first set of a new generation of models of massive stars of solar composition extending between 13 and 120 \msun, computed with and without the effects of rotation. We included two instabilities induced by rotation, namely the meridional circulation and the shear instability. We implemented two alternative schemes to treat the transport of the angular momentum: the advection-diffusion formalism and the simpler purely diffusive one. The full evolution from the Pre Main Sequence up to the presupernova stage is followed in detail with a very extended nuclear network. The explosive yields are provided for a variety of possible mass cut and are available at the website \url{http://www.iasf-roma.inaf.it/orfeo/public{\_}html}.

We find that both the He and the CO core masses are larger than those of their non rotating counterparts. Also the C abundance left by the He burning is lower than in the non rotating case, especially for stars of initial mass 13-25 \msun, and this affects the final Mass-Radius relation, basically the final binding energy, at the presupernova stage. The elemental yields produced by a generation of stars rotating initially at 300 km/s do not change substantially with respect to those produced by a generation of non rotating massive stars, the main differences being a slight overproduction of the weak {\it s}-component and a larger production of F. Since rotation also affects the mass loss rate, either directly and indirectly, we find substantial differences in the lifetimes as O-type and WR-subtypes between rotating and non rotating models. The maximum mass exploding as type IIP supernova ranges between 15 and 20\msun in both sets of models (this value depending basically on the larger mass loss rates in the Red Super Giant phase due to the inclusion of the dust driven wind). This limiting value is in remarkable good agreement with current estimates.

\end{abstract} 

\section{Introduction}

The life of an isolated star is univocally determined once the initial mass, chemical composition, distribution of the angular momentum and the possible presence of magnetic fields are fixed. For historical reasons the so called "classical" models are, or at least were, computed neglecting the presence of rotation and magnetic fields. The comparison between these "classical" models and their observational counterparts showed over the years many discrepancies that cannot be explained by these models. Among the others, the width of the upper MS and the color distribution of massive stars in the HR diagram \citep[][and references therein]{massey2003}, the relative proportions between O stars and the various WR subtypes \citep[][and references therein]{mm03}, the surface enhancement of He and N in MS stars \citep{mokiem06,mokiem07,untetal09}, the ratio of blue to red supergiants as a function of the metallicity \citep{lm95}, the peculiar chemical composition of many very metal poor stars \citep[][and references therin]{lc12}. In order to improve the comparison between models and observed stars, uncertain physical phenomena like the extension of the convective regions and the efficiency of mass loss were varied more or less arbitrarily. In this way it was possible to improve one discrepancy or the other, but no satisfactory general scenario has ever emerged from all these tests.

Rotation in principle could be a good candidate to reduce all together these discrepancies since it may influence the mass loss rate as well as trigger flows of matter within the star. Though the possible influence of rotation on the evolution of a star was already outlined even before the second world war \citep{vz24,eddington29,gratton45}, only in the last decade or so extended sets of massive star models computed taking into account rotation began to appear. 

After the pioneering works by \citet{kt70}, \citet{es76}, \citet{pin89} and \citet{kw90}, 
grids of rotating massive stars have been computed and discussed by
\citet{mm00,mm03,hirschi04,mm05,decressin07,ekstroem08,ekstroemetal12} and
\citet{hlw00,hl00,hws05,yl05,yln06,brott11}.

More than a decade ago we started a project aimed to study and progressively refine both the evolutionary properties of massive stars from the MS to the core collapse and the passage of the shock wave through the mantle in order to quantitatively determine the final explosive yields of a large number of nuclear species. In this paper we discuss the inclusion of rotation in our stellar evolutionary code, the FRANEC, and we present a first set of solar metallicity massive star models evolved up to the core collapse together to the following explosion and the final explosive yields. In a companion paper (in advanced preparation) we will provide a full database of tracks extending in metallicity between z=0 and z=solar. The paper is organized as follows: section 2 is devoted to the implementation of rotation in the FRANEC code while the calibration of the rotationally induced instabilities is discussed in section 3. Section 4 and 5 are devoted, respectively, to the H and He burning phases. The advanced burnings and the explosive yields are discussed in sections 6 and 7. A final summary and conclusions follows.

\section{Set up of the FRANEC}\label{franec}

The latest version of the FRANEC (Frascati RAphson Newton Evolutionary Code), major release 6, now includes the effects of rotation on the evolution of a star.  Here we describe only the main features of this new release with respect to the previous one, which is described in detail in \cite{lc03}.

In principle the inclusion of rotation in an evolutionary code would require at least a 2D scheme (if not a 3D one). \cite{kt70} and \cite{es76} showed, however, that under few proper assumptions it is possible to simulate the mechanical and thermal distortions induced by rotation in a 1D code. In the following we will not repeat the detailed derivation of the basic equations in presence of rotation, but we will simply summarize the basics of their derivation and address few comments we consider of interest to properly understand the differences among the various approaches followed by different groups.  We strongly suggest the reader to read carefully the works of \cite{kt70} and \cite{es76} for a detailed derivation of the equations in presence of rotation. 

The basic assumptions under which it is possible to simulate the average mechanical and thermal distortions induced by rotation in a 1D code, following the scheme proposed by \cite{kt70}, are:

\begin{itemize}
\item cylindrical symmetry for the angular velocity $\omega$ (needed in order to define equipotential surfaces and to keep
the equations for the stellar structure in the same form as in spherical symmetry except for the introduction of two "form factors" that take into account the mechanical and thermal distortions induced by rotation);

\item $\omega={\rm const}$ along the equipotentials and Roche approximation 
(needed in order to compute easily the shape of the equipotentials).

\end{itemize}

Let us note, at this point, that this last assumption initially chosen for sake of simplicity, received additional support from the work of \cite{z92}, who showed that the onset of a strong horizontal turbulence allows the growth of vigorous mixing that very probably keeps the chemical composition and the angular velocity constant along the isobars. Such an approximation is currently referred to as "shellular rotation". Since one of the consequences of the first assumption (cylindrical symmetry for the angular velocity $\omega$) implies that the equipotentials coincide with the isobars, let us therefore stress here that the \cite{kt70} scheme is based on the shellular rotation approximation even though it was not named in this way at that time. 

As already mentioned above, the combination of cylindrical symmetry and shellular rotation necessarily implies a solid body rotation and this would limit the application of the \cite{kt70} scheme to differentially rotating stars.

A strong support to the work of \cite{kt70} came a few years later when \cite{mm97} showed that it was possible to maintain the equations of the stellar structure as modified by \cite{kt70} even dropping the requirement of cylindrical symmetry (and hence the existence of a conservative field). The price necessary to pay in this case is that the unique $\rho$ and T that can be associated to an isobar in the conservative case must now be replaced by appropriate averages on each isobar \citep[see][for more details]{mm97}. Since the conservative and non conservative case differ basically for the meaning of the variables (T and $\rho$ constant on an isobar in the conservative case and just average values in the non conservative case), from any practical point of view, the same scheme can be interpreted in either of the two ways.

In addition to the distortions induced by rotation on the structure of a star, it is very important to consider the physical phenomena that may lead to a significant redistribution of the angular momentum and/or chemical composition. In addition to the more traditional dynamical instabilities (determined on the basis of the Scwharzchild and/or Ledoux criteria), including semiconvection in H burning, induced semiconvection in He burning and the possible occurrence of some overshooting, we have also implemented two rotational instabilities: meridional circulation and shear instability. 

Once the instabilities to be included in the code have been identified or, actually, chosen, it is crucial to understand which is the equation that controls the mixing of both the angular momentum and the chemicals. As for the transport of the angular momentum, there are basically two different approaches. The first one, more rigorous, takes into account the fact that the equation that controls the transport of the angular momentum is not a pure diffusive equation but it has an advective term, which means that the angular momentum is not necessarily spreaded out, flattened, but that it could be even steepened by the redistribution. We implemented in the FRANEC the same formulation adopted by \cite{tal97} and firstly derived by \cite{cz92}:

\begin{equation}
\rho r^{2}{dr^{2}\omega \over dt}={1 \over 5}{\partial \over \partial r}(\rho r^{4}U\omega )+
{\partial \over \partial r}\left(\rho r^{4}D_{s.i.}{\partial \omega \over \partial r}\right)
\end{equation}

where U represents the radial component of the velocity of the meridional circulation, $D_{s.i.}$ the diffusion coefficient corresponding to the shear instability and all other quantities have their usual meaning. We adopted the formulation for U given by \cite{mz98} (their eq. 4.38) and the expression of $D_{s.i.}$ proposed
by \cite{tz97} and \cite{pal03}:

\begin{equation}
D_{s.i.}=\frac{8}{5}\frac{R_{\mathit{ic}}(rd\omega/\mathit{dr})^{2}}{N_{T}^{2}/(K+D_{h})+N_{\mu }^{2}/D_{h}}
\label{dshear}
\end{equation}

where $N_{T}^{2}=\frac{g\delta }{H_{P}}(\nabla _{\mathit{ad}}-\nabla_{\mathit{rad}})$,
$N_{\mu }^{2}=\frac{g\delta }{H_{P}}(\frac{\varphi}{\delta }\nabla _{\mu })$ and
$R_{\mathit{ic}}=\frac{1}{4}$. $\rm D_h$ is the coefficient of horizontal turbulence given by $D_h \simeq | r~U |$ \citep{z92}, while $K=\frac{4acT^3}{3C_pk\rho^2}$ is the thermal diffusivity \citep{mm00}.

Alternative expressions for $D_h$ were proposed by \cite{m03} and \cite{mpz04} (both leading to a larger efficiency of the horizontal turbulence), but we preferred to adopt the one proposed by \cite{z92} in this (our) first paper on rotation for a better compatibility with the set of models published by \cite{mm03}. For the same reason we did not adopt the alternative expression proposed by \cite{m97} for the diffusion coefficient of the shear mixing.  

As for the transport of the chemicals, \cite{cz92} showed that in the asymptotic regime the
transport of the chemical composition within a radiative region due to rotation can be described by a pure diffusive process. In this case the diffusion coefficient is given by

\begin{equation}
D = D_{s.i.}+D_{m.c.}
\label{diffcoeff}
\end{equation}

where $D_{s.i.}$ is given by equation \ref{dshear} while $D_{m.c.}$ is

\begin{equation}
D_{m.c.} \simeq {| r~U |^2 \over 30 D_h}
\label{diffcoeffmc}
\end{equation}

as suggested by \cite{z92}.

Since the numerical implementation of the advective-diffusive equation for the transport of the angular momentum is quite challenging and since the current computations show that in most cases the redistribution of angular momentum goes in the direction of flattening the $\omega$ profile, the formally correct equation is often substituted by a much simpler diffusion equation. We implemented also this alternative approach in the code. For the pure diffusion case we used a standard diffusion equation with the diffusion coefficients given by 
equations \ref{dshear}, \ref{diffcoeff} and \ref{diffcoeffmc}.
The velocity $U$ that enters in equation \ref{diffcoeffmc} now is not the more general one provided by \cite{mz98} (their eq. 4.38) but the simpler expression provided by \cite{kw90}:

\begin{equation}
U=\frac{8}{3}\frac{\omega^2 r}{g}\frac{L}{M g}\frac{\gamma-1}{\gamma}\frac{1}{\nabla_{ad}-\nabla} \left( 1-\frac{\omega^2}{2 \pi G \rho} \right)
\label{velmckipp}
\end{equation}

When the pure diffusion approximation is adopted, the diffusion coefficients used for the mixing of the chemical composition are the same adopted for the transport of the angular momentum (equations \ref{dshear}, \ref{diffcoeff}, \ref{diffcoeffmc} and \ref{velmckipp}).

The set of equations formed by the four ones describing the physical structure of the star plus the "n" ones describing the total chemical evolution (i.e. the local burning due to the nuclear reactions plus the various kinds of mixing: convection, semiconvection and rotationally induced mixing) are coupled together and solved simultaneously  by means of a relaxation technique. The temporal evolution of the angular momentum together to the determination of the profile of the velocity of the meridional circulation is solved separately, again by means of a relaxation technique. 

Mass loss has been included following the prescriptions of \cite{val00,val01} for the blue supergiant phase ($\rm T_{eff}>12000~K$), \citet{dejager88} for the red supergiant phase ($\rm T_{eff}<12000~K$) and \citet{nl00} for the Wolf-Rayet phase. The criteria adopted to define the different WR subclasses are the same adopted in \cite{lc06}. The enhancement of the mass loss due to the formation of dust during the red supergiant phase has been included following the prescriptions of \citet{vanloonetal05}. Mass loss is enhanced, in rotating models, following the prescription of \cite{hlw00}.

The nuclear network includes 163 isotopes (from H to $\rm ^{\rm 97}Mo$) and 448 reactions for H and He burning, and 282 isotopes (from H to $\rm ^{\rm 98}Mo$) and 2928 reactions for the more advanced nuclear burning stages up to the procollapse phase. In total 293 isotopes and about 3000 processes were explicitly included in the various nuclear burning stages. The database of the cross sections adopted in this first paper on rotating stellar models is the same adopted in \cite{lc06}.

The heavy element solar mixture is the one provided by \citet{agss09}. The initial values of the global metallicity $Z$ and of the He abundance $Y$ have been set to
$Z=1.345 \times 10^{-2}$ and $Y=0.265$, respectively. These values have been obtained by requiring that a $1~{\rm M_\odot}$, computed by taking into account the
atomic diffusion, reproduces the main properties of the present Sun (radius, luminosity, rotation rate and helium abundance) after ${\rm t_\odot}=4.57~{\rm Gyr}$
\citep[see][for more details]{scl97}.

All models are started on the Hayashi track when the central temperature is of the order of $10^5$ K. The angular velocity is raised above zero when the stellar model reaches the Main Sequence: at this point $\omega$ is raised slowly and kept constant throughout the star until the chosen surface equatorial velocity is reached. Less than 1\% of H is burnt in this phase.

\section{Calibration of the mixing efficiency}\label{calibration}

The diffusion coefficients discussed so far are intrinsically uncertain and must be regarded as an estimate of the efficiency of the transport of both the angular momentum and the chemicals. For this reason we decided to consider two free parameters, namely, $f_c$ and $f_\mu$, to be calibrated somehow. Similar parameters have already been introduced by, e.g., \cite{pin89}, \cite{hlw00} and \cite{brott11}. The first free parameter, $f_c$, simply multiplies the total diffusion coefficient defined in equation \ref{diffcoeff} and adopted to transport the chemicals, i.e.:

\begin{equation}
D = f_c \times (D_{s.i.}+D_{m.c.})
\end{equation}

This means that $f_c$ directly controls the speed at which the chemical composition is mixed in the various zones inside a star without affecting the transport of the angular momentum. The second free parameter, $f_\mu$, multiplies the gradient of molecular weight ($\nabla_\mu^{\rm adopted}=f_\mu \times \nabla_\mu$) and regulates the influence of this quantity on the mixing of both the angular momentum and the chemical composition. This parameter is necessary because the inclusion of the gradient of the molecular weight at its face value strongly inhibits the transport \citep{pin89,m97,mm97,tz97}.

\begin{figure}
\epsscale{.99}
\plotone{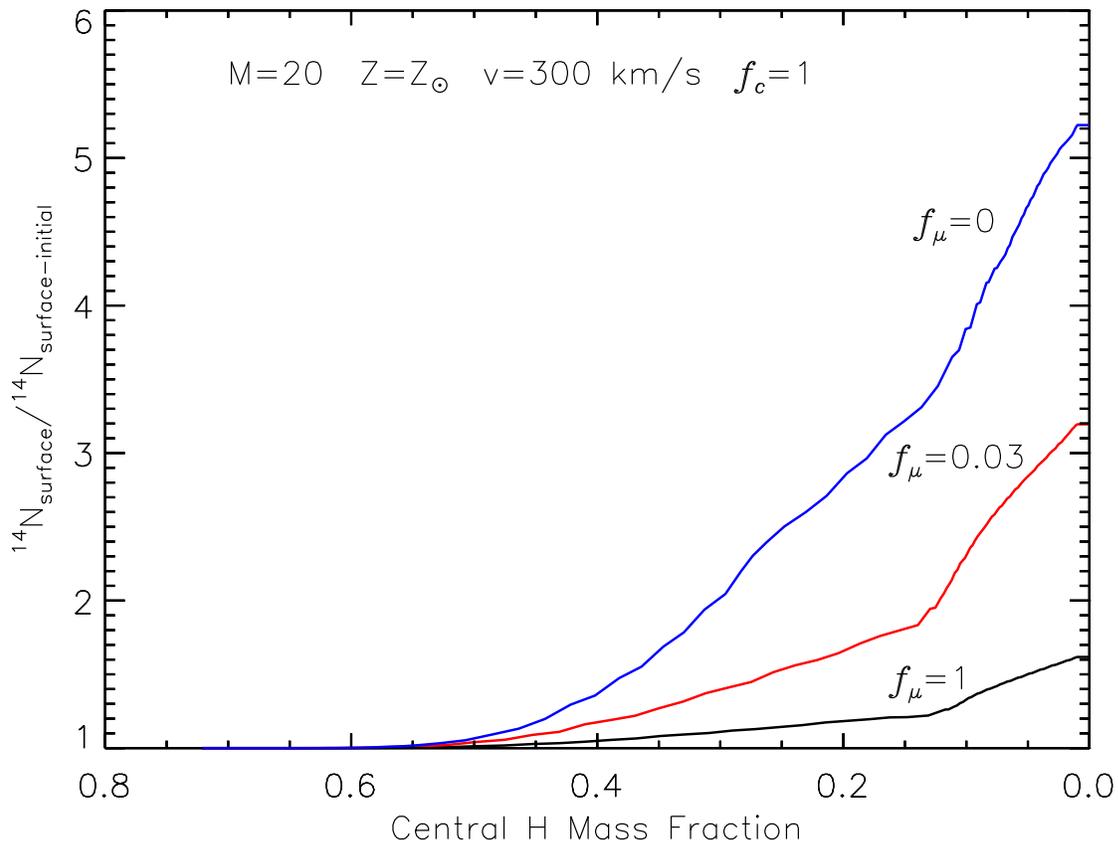}
\caption{Surface $\rm ^{14}N$ abundance, normalized to the initial value, as a function of
the central H mass fraction for three models of 20 \msun having an initial solar composition and an
initial equatorial velocity of 300 km/s.
All the models are computed with $f_c=1$ but for three different values of $f_\mu$, namely, $f_\mu=0~{\rm (lower,~black~line)}$, $~ 0.03~{\rm (middle,~red~line)}$ and $1~{\rm (upper~blue~line)}$.}
\label{checka03b}
\end{figure}

In the present set of models, computed by adopting the advection-diffusion scheme, we chose a conservative approach by fixing $f_c=1$ and by calibrating $f_\mu$ so that stars of solar metallicity in the mass range 15-20 $\rm M_\odot$, and settling on the main sequence with an initial equatorial velocity of 300 km/s, increase the initial surface nitrogen abundance by roughly a factor of three at core H depletion \citep[see, e.g.,][]{hlw00}. 
The $f_\mu$ value that gives such an increase is 0.03. Figure \ref{checka03b} shows the run of the surface N abundance as a function of the central H mass fraction for three models of 20 $\rm M_\odot$, all computed with $f_c=1$ but for three different values of $f_\mu$, namely, $f_\mu=0,~0.03,~1$. The largest nitrogen increase is obviously obtained for the case $f_\mu=0$ and is of the order of 5.

\begin{figure}
\epsscale{.99}
\plotone{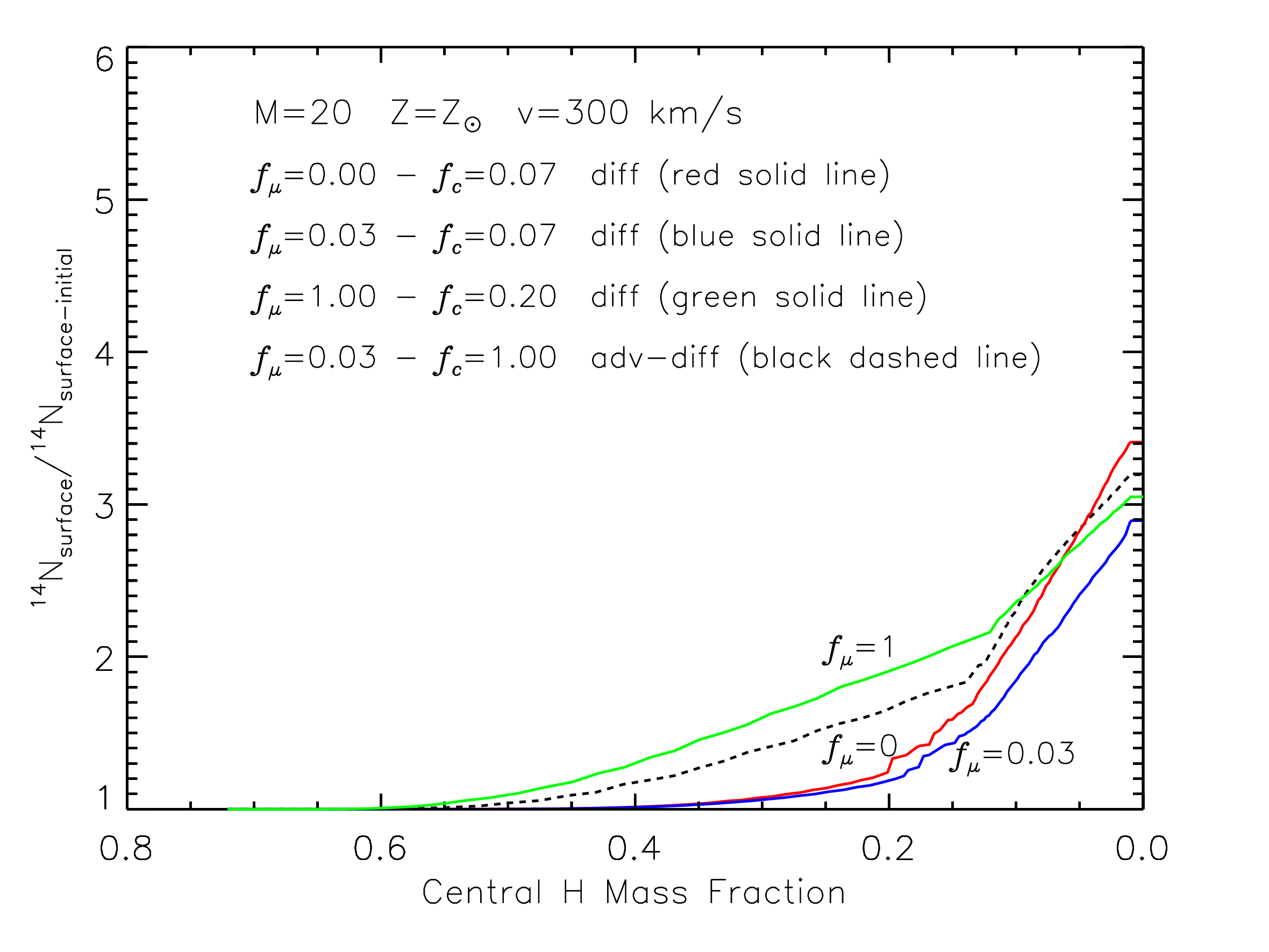}
\caption{Surface $\rm ^{14}N$ abundance, normalized to the initial value, as a function of
the central H mass fraction for four models of 20 \msun having an initial solar composition and an
initial equatorial velocity of 300 km/s, computed with different choices of the angular momentum transport
scheme and of the $f_c$ and $f_\mu$ parameters, respectively: a) advection-diffusion, $f_\mu=0.03, f_c=1$ (black dashed line, the reference model); b) pure diffusion, $f_\mu=0,~f_c=0.07$ (red solid line); c) pure diffusion, $f_\mu=0.03,~f_c=0.07$ (blue solid line); d) pure diffusion, $f_\mu=1,~f_c=0.2$ (green solid line).}
\label{checka04b}
\end{figure}

How already stated above, we implemented also a pure diffusive scheme to treat the transport of the angular momentum. Also this scheme obviously requires a calibration of the two parameters $f_c$ and $f_\mu$. In this case it is not possible to use $f_c=1$ because the mixing is now much more efficient. Since there is more than one combination of $f_c$ and $f_\mu$ values that may provide the same surface N enhancement, we explored some alternatives by calibrating the value of $f_c$ for three values of $f_\mu$, namely $f_\mu=0,~ 0.03,~1$ (the same values shown in Figure \ref{checka03b}). The values of the $f_c$ parameter that provide
the requested surface nitrogen enrichment in each one of the three 20 \msun models are $f_c=0.07,~0.07,~0.2$, respectively, and the corresponding surface N enhancements are shown in Figure \ref{checka04b}. Obviously the larger the $\mu$ barrier the larger the $f_c$ value needed to obtain the same mixing efficiency. The black dashed line in Figure \ref{checka04b} refers to our "standard" case, i.e. the model computed in the advective-diffusive framework with $f_c=1$ and $f_\mu=0.03$.

It is worth noting that a surface N enhancement of the order of 3 at core H depletion, for solar metallicity rotating stars in the mass range 15-20 \msun with $v_{ini}\simeq 300~{\rm km/s}$, is comparable to what is obtained also by other groups. Without having the aim of discussing their calibrations, let us simply note that, for example, \cite{mm03} obtain $\rm N_{H=0}/N_{ini}=2.9$ for a 
20 \msun star with $\rm Y_{ini}=0.275,~Z_{ini}=0.02~{\rm and}~v_{ini}=300~{\rm km/s}$, while
\cite{brott11} find $\rm N_{H=0}/N_{ini}=2.5$
for a 20 \msun star with $\rm Y_{ini}=0.2638,~Z_{ini}=0.008~{\rm and}~v_{ini}=274~{\rm km/s}$.

A last point we want to stress here concerns the fact that the surface N enhancement ($\Delta$N) at core H depletion  depends significantly also on the initial abundances of Carbon and Nitrogen. The reason is straightforward, since the $\Delta$N comes form the conversion of C into N (CN cycle) that occurs in the deep interior of the star: the smaller the initial N abundance the larger the $\Delta$N, and also the larger the initial C abundance the larger the $\Delta$N as well. Such an occurrence has been well shown quantitatively by \cite{brott08}.

As for the size of the convective regions, we did not attempt any calibration but we included an overshooting of 0.2 $\rm H_p$ for compatibility with our previous set of models \citep{lc06}. 

\section{Core H Burning}

The core H burning phase is by far the longer lasting evolutionary phase of a star.
Therefore it is either the phase during which a star can
be more probably observed and also the one during which the secularly unstable rotationally mixing phenomena have more time to operate. How we have already stated in the previous sections, the influence of rotation on the evolution of a star 
during core H and core He burning phases has been already discussed many times. 
Hence here we will only discuss a few aspects we consider of interest for the reader. 

\begin{figure}
\epsscale{.99}
\plotone{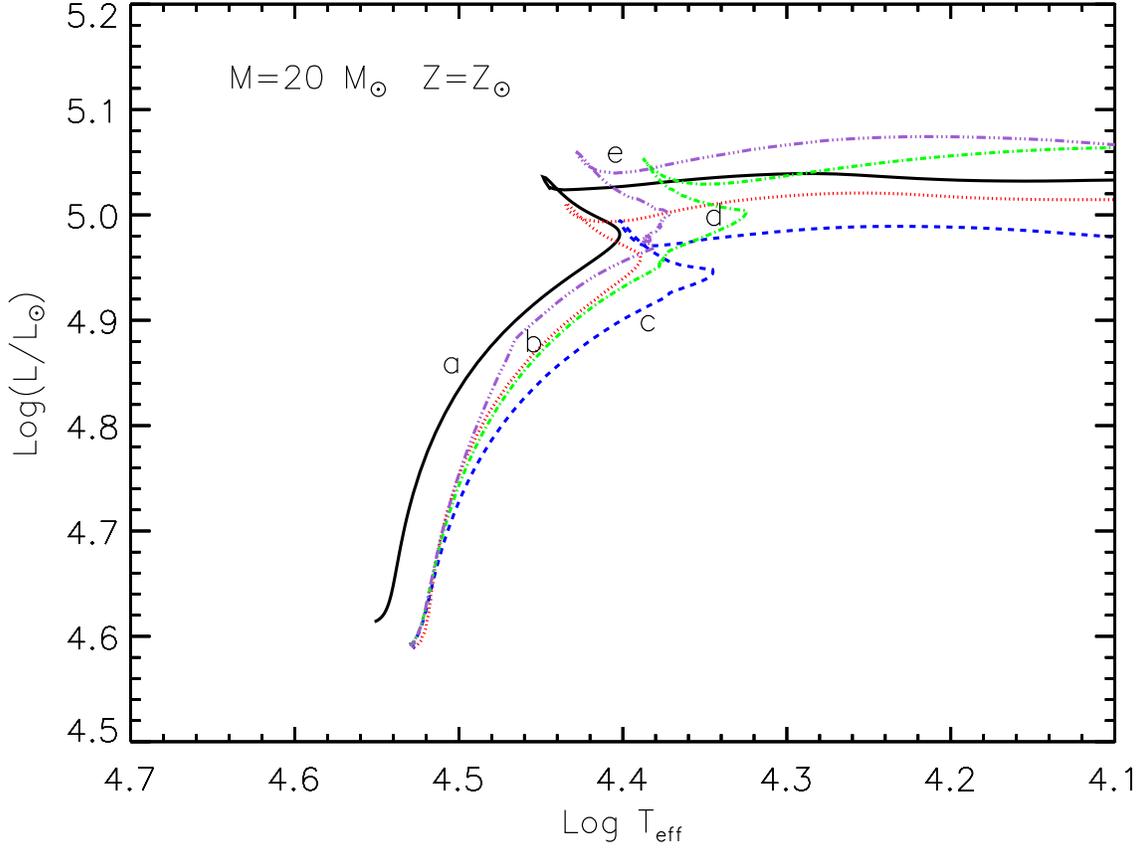}
\caption{Evolutionary tracks of a 20 \msun model of solar composition during core H burning, computed with different choices
on the inclusion of rotation: 
a) no rotation (black solid line); 
b) only the mechanical and thermal distortion induced by rotation have been taken into account, i.e.,
$f_P$ and $f_T$, without any kind of rotationally induced mixing (red dotted line); 
c) like model b) but including also the transport of the angular momentum by either the meridional circulation and the shear instability
(blue dashed line); 
d) like model c) but adding also the mixing of the chemical composition due to the meridional circulation (green dash-dot line); 
e) model computed by taking into account all the effects of rotation, i.e., mechanical and thermal distortion, transport
of angular momentum and chemical species due to both meridional circulation and shear instabilities (violet dash-dot-dot line).}
\label{checka02b}
\end{figure}

Let us firstly show how the various effects of rotation sum up to determine the position of a rotating star on the Main Sequence and affect its core H burning phase. Of course, the influence of rotation will depend first of all on the amount of angular momentum injected in the star at the beginning: all models computed for this test have an initial surface equatorial velocity of 300 km/s, which corresponds (in these specific models!) to an $\omega / \omega_{\rm crit}\sim 0.6$,
where $\omega_{\rm crit}$ is the classical critical angular velocity and, in the framework of the Roche model, is given by $\omega_{\rm crit}=(2/3)^{3/2}(GM/R_{pb}^3)^{1/2}$, where $R_{pb}$ is the polar radius of the star when the surface rotates with the critical velocity. 
This velocity is already larger than the bulk of observational data available up to now. Figure \ref{checka02b} shows five tracks: the first one (black solid line) refers to a standard 20 \msun computed without rotation while the second (red dotted line) refers to the evolution of the same stellar model in which only the $f_P$ and $f_T$ factors have been taken into account. This means that only the mechanical and thermal distortion of the structure is considered without any kind of rotationally induced mixing: it goes without saying that this automatically implies local conservation of the angular momentum. The red dotted line is shifted towards slightly lower luminosities and surface temperatures as a consequence of the centrifugal force that sums to the pressure gradient in sustaining the star (i.e., the effective gravity is reduced and the star is more expanded). The third track (blue dashed line) refers to a model computed like the red one but including also the transport of the angular momentum by either the meridional circulation and the shear instability. The inclusion of the transport of the angular momentum increases progressively the shift of the model towards lower effective temperatures because it raises the centrifugal force in the mantle of the star (hence lowers even more the effective gravity). The green dash-dot line refers to a model computed by adding also the mixing of the chemical composition due to the meridional circulation: in this case the star shifts slightly blueward during the core H burning phase because the partial mixing of the matter throughout the whole star smears out the gradient of molecular weight and makes the star more compact on average
(see below). Note that this mixing increases the size of the H depleted zone (in some sense it acts similarly to the overshooting) and this is clearly shown by the much higher luminosity at which the overall contraction occurs. The last sequence (violet dash-dot-dot line) refers to the complete model in which also the shear instability contributes to the mixing of the matter. Since the net result is simply a more robust mixing of the matter inside the star, this stellar model shows a more pronounced blueward shift and a higher luminosity at the end of the central H burning.

\begin{figure}
\epsscale{.83}
\plotone{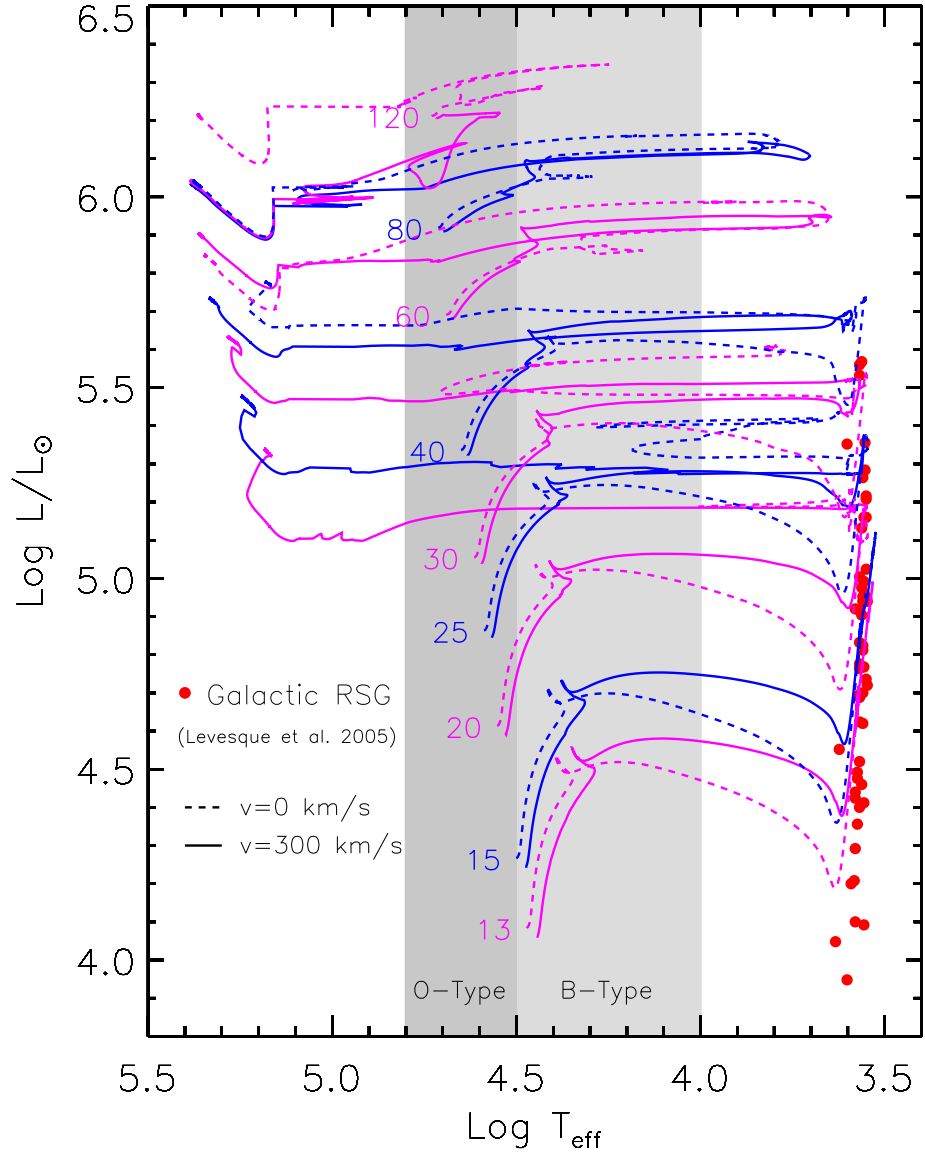}
\caption{Evolutionary path in the HR diagram of rotating (solid lines) and non rotating (dashed lines) models. Also shown in Figure is
the location of the Red Supergiant stars in the Galaxy \citep{levesque05}}
\label{hrtot}
\end{figure}

A basic comparison between the H burning phase of two different sets of models (in this specific case rotating versus non-rotating models) naturally implies  a comparison of the paths in the HR diagram, of the internal structures, of the lifetimes and of the total amount of mass lost. Also possible differences in the surface chemical composition are extremely relevant first of all because the surface of a star is our basic observational counterpart. In the case of rotating models also the variation of the surface velocity with time must be discussed because this is another quantity that can be observed.

The evolutionary path of rotating (with an initial equatorial velocity of $v_{ini}=300~{\rm km/s}$) and non rotating models in the HR diagram is shown in Figure \ref{hrtot} while the main evolutionary properties of the two sets of models are reported in Tables 1 and 2. The various columns report, for each mass: 1) the evolutionary phase, 2) the lifetime (in yr), 3) the mass size of the convective core (in solar masses), 4) and 5) the average effective temperature and luminosity, 6), 7) and 8) the total mass, the He core mass and the CO core mass (in solar masses), 9) the equatorial velocity (in km/s), 10) the surface angular velocity (in $s^{-1}$), 11) the ratio of the surface angular velocity with respect to the critical angular velocity, 12) the total angular momentum (in $10^{53}~g~cm^2~s^{-1}$), 13), 14) and 15) the surface abundances of H, He and N in mass fraction, 16) and 17) the N/C and N/O surface ratios. According to what was discussed above, all rotating models settle on a MS position slightly less luminous and redder than their respective non rotating models. The path of the models in the HR diagram will depend on the balance between the outward transport of the angular momentum (that tends to push the models towards lower effective temperatures) and the transport of the chemicals (that, viceversa, tends to push the track blueward because of the smearing out of the molecular weight gradient). The transport of the angular momentum prevails in the lower mass stars hence the tracks of these models remain roughly parallel to those of the non rotating ones. On the contrary, the transport of the chemicals prevails in the higher mass stars so that the path of these models tends to shorten and even cross that of their respective non rotating counterparts. It is therefore clear that the larger the efficiency of mixing of the chemicals the smaller the minimum "crossing mass", i.e. the mass for which the rotating and non rotating tracks intersect during the H burning phase. An inspection of the sets of models published by \cite{ekstroemetal12} and \cite{brott11} shows that in both bases such minimum "crossing mass" is much lower than in our case. This clearly points towards a much more efficient mixing in their models. This is well confirmed by Figure \ref{confazoto} (see below) that shows a comparison of the surface N enrichment among the \cite{ekstroemetal12}, \cite{brott11} and our models: the \cite{ekstroemetal12} model shows a very efficient mixing since the beginning that prolongs for the whole H burning phase while the \cite{brott11} one shows an efficient mixing for more the first 40\% of the H burning phase or so. Our model, viceversa, show a rapid growth of the surface N abundance only towards the end of the central H burning phase. It is not easy to understand where these differences come from because each set of models assumes different coefficients for the mixing of the chemicals: we adopt the diffusion coefficient proposed by \cite{tal97}, that includes the contribution of the horizontal turbulence as given by \cite{z92}, while \cite{ekstroemetal12} adopt the coefficient provided by \cite{m97} that does not include the horizontal turbulence. \cite{brott11} adopt a even different approach \citep{hlw00} and moreover take into account also the transport of the angular momentum by magnetic fields (but not the transport of the chemicals). 

The very strong dependence of the mass loss rate (in MS) on the luminosity (i.e. on the initial mass) naturally divides massive stars in two subgroups: a first group formed by those stars that do not lose a significant amount of mass during core H burning ($M \lesssim 30~{\rm M_\odot}$) and those whose MS evolution is dominated by mass loss.

\begin{figure}
\epsscale{.99}
\plotone{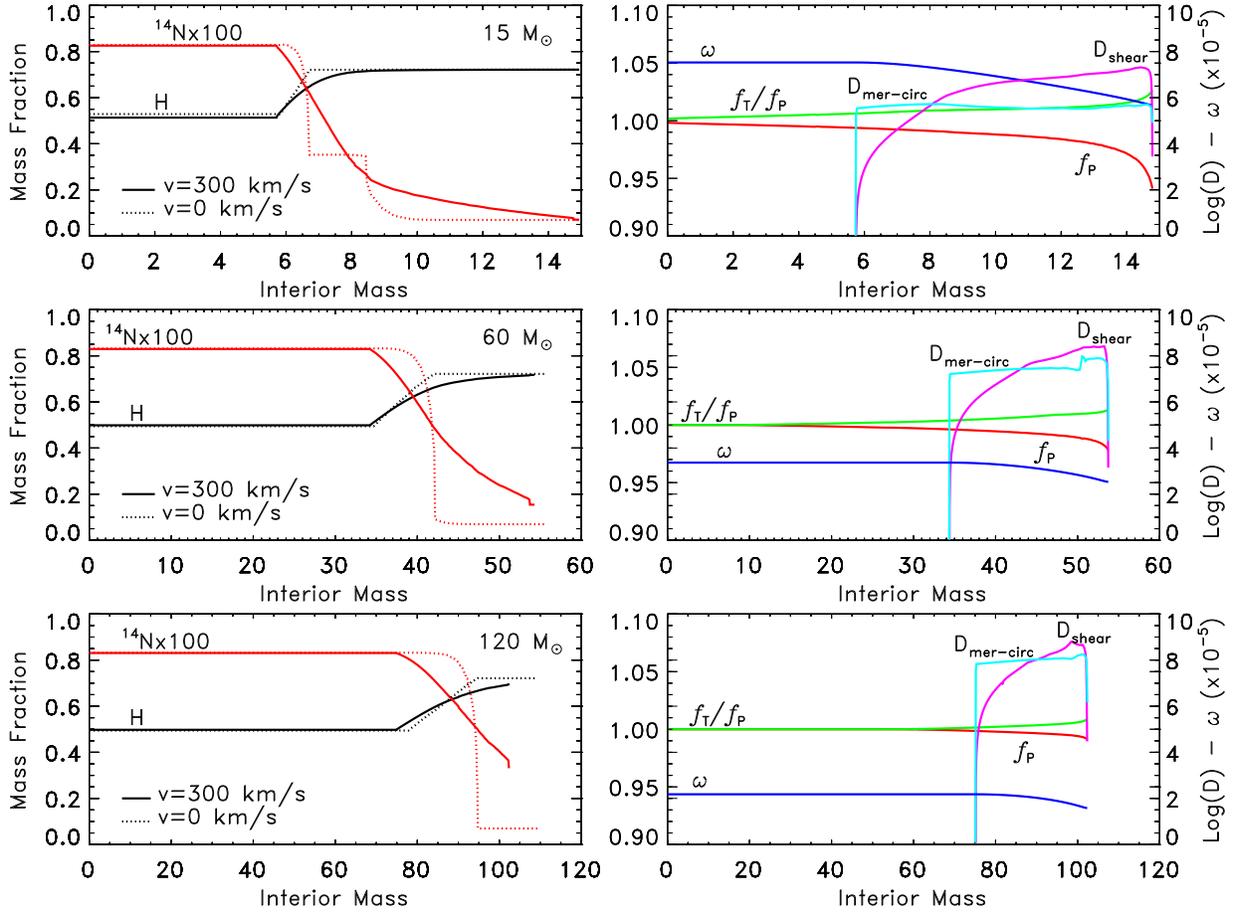}
\caption{Internal chemical composition (left panels) and physical properties of rotating models (right panels) at the time the central H mass fraction
has decreased to $\sim 0.5$ during core H burning, for three selected models, i.e., 15, 60 and 120 \msun. In the left panels it is shown the
internal profiles of H (black line) and \nuk{N}{14} (red line) mass fractions - the solid and dotted lines refer to the rotating and non rotating models, respectively. Note that in order to improve the readability of the figure the \nuk{N}{14} mass fraction has been multiplied by a factor of 100.
The right panels provide some information about the interior properties of the rotating models: the angular velocity $\omega$ (blue line); 
the diffusion coefficients driving the transport of the chemical species [shear instability (magenta line) and meridional circulation (cyan line)];
the two "form factors" $f_P$ (red line) and $f_T/f_P$ (green line).}
\label{confstru1}
\end{figure}

\begin{figure}
\epsscale{.99}
\plotone{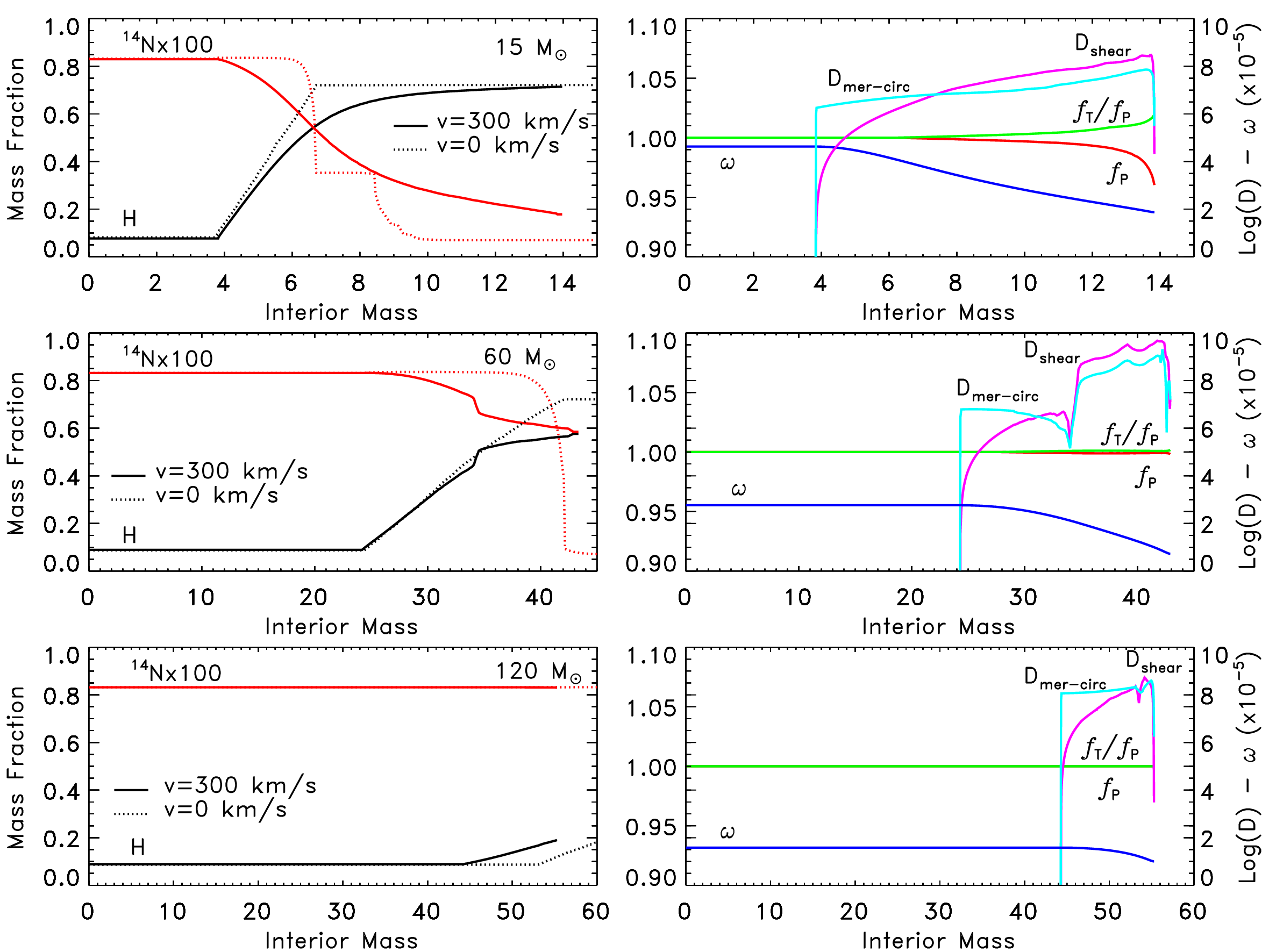}
\caption{Same as Figure \ref{confstru1} but at the time the central H mass fraction has decreased to $\sim 0.1$ during core H burning.}
\label{confstru2}
\end{figure}

\begin{figure}
\epsscale{.99}
\plotone{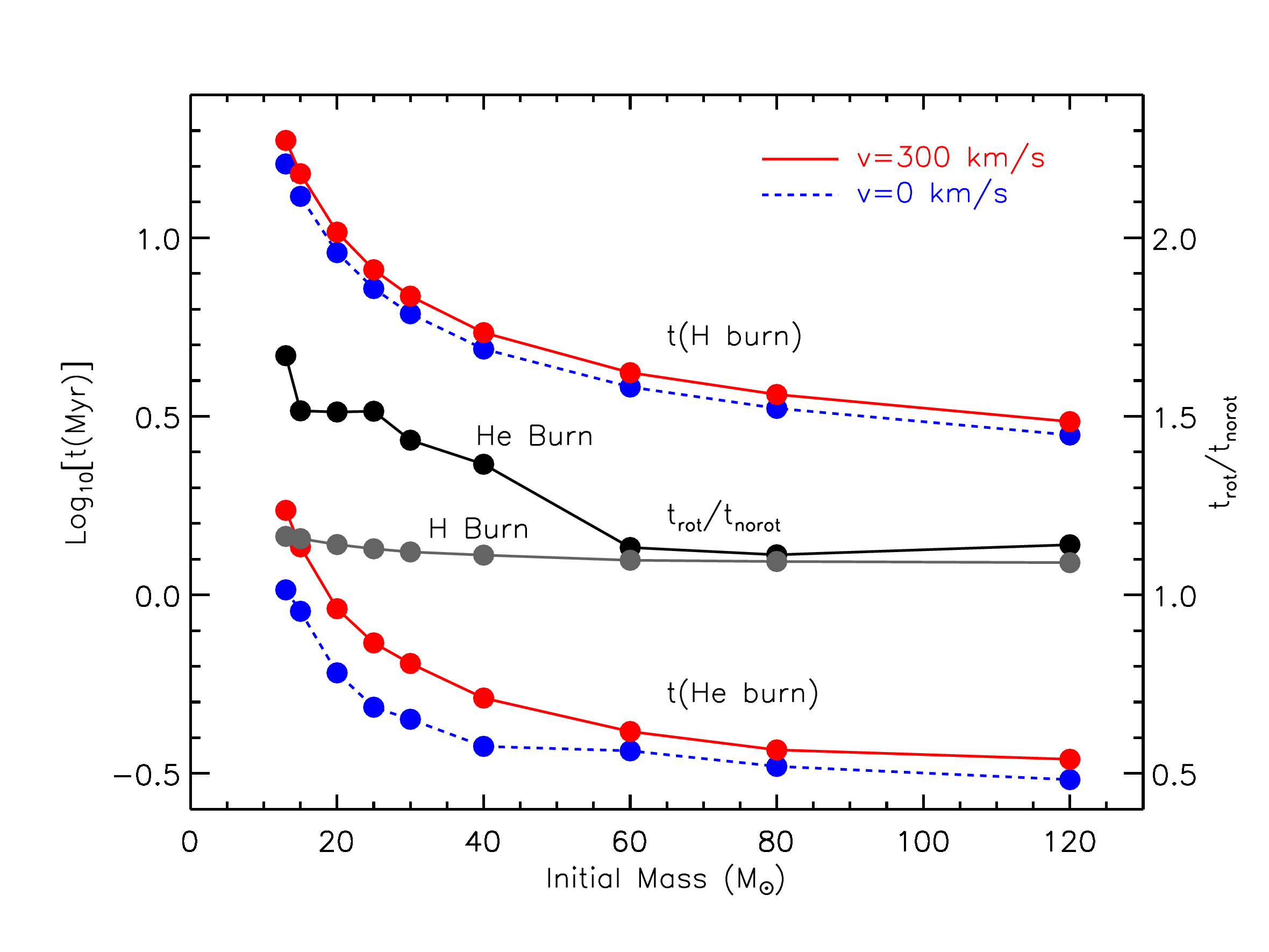}
\caption{H and He burning lifetimes (left axis) in rotating (red solid lines) and non rotating (blue dashed lines) models. 
The ratios of the corresponding H-burning and He-burning lifetimes obtained in rotating and non rotating models is reported in the right axis.}
\label{times}
\end{figure}

During core H burning, rotationally induced mixing drives a continuous slow ingestion of fresh fuel into the convective core, as well as a slow mixing of the freshly synthesized H burning products into the envelope of the star. Such a mixing is clearly visible by comparing both the H and the \nuk{N}{14} profiles, in rotating and non rotating models, at different stages during core H burning (Figures \ref{confstru1} and \ref{confstru2}). In particular, it is readily evident in Figures \ref{confstru1} and \ref{confstru2}, that the chemical profiles are much shallower in rotating models than in the non rotating ones. This is a clear effect of a slow diffusive mixing between the H convective core and the radiative H-rich envelope. Such a mixing is driven by both the meridional circulation, that dominates at the base of the radiative envelope, and by the shear instability that is more efficient in the outer layers (see right panels of Figures \ref{confstru1} and \ref{confstru2}). The main effect of this slow ingestion of fresh fuel within the H convective core is that of slightly increasing the H burning lifetime of rotating models with respect to the non rotating ones. The differences, however, do not exceed the 15\% at the lower end of the mass interval (13 \msun) and flatten out at a 10\% level above 60 \msun or so (see Figure \ref{times}). Another interesting information provided by Figures \ref{confstru1} and \ref{confstru2}, and in particular from the internal profiles of the two "form factors" $f_P$ and $f_T/f_P$, is that the departure from the spherical symmetry (i.e. where $f_P$ significantly decreases below 1.0) is negligible in the more internal zones while it progressively increases moving outward in mass. 

\begin{figure}
\epsscale{.99}
\plotone{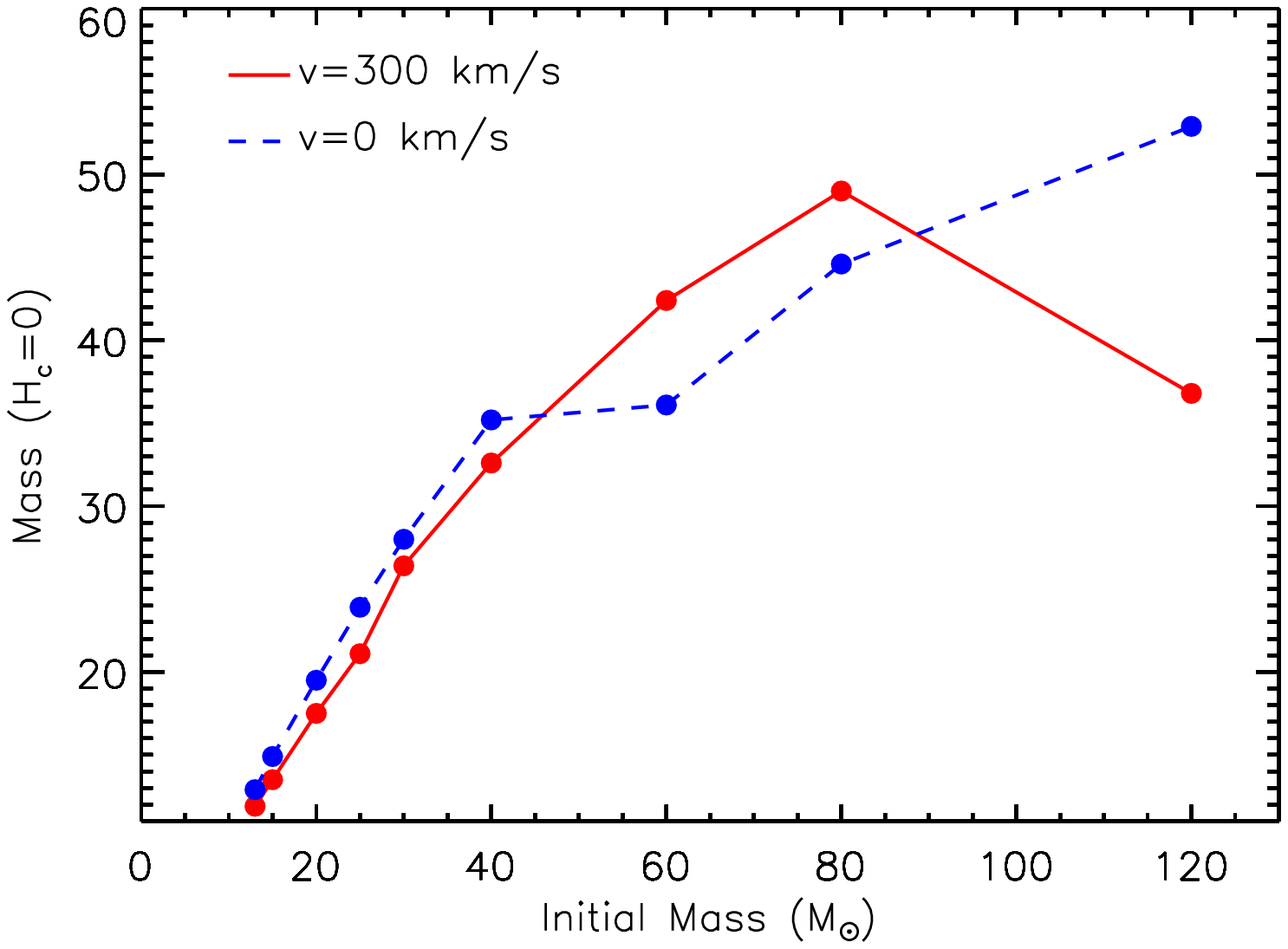}
\caption{Total mass at core H depletion for rotating (red solid line) and non rotating (blue dashed line) models.}
\label{massHburn}
\end{figure}

The different paths followed by rotating and non rotating models in the HR diagram together to the slight differences in the lifetime naturally lead to different mass loss histories and hence to different total masses at core H exhaustion. Figure \ref{massHburn} shows such a comparison. Since rotating models have lower effective temperatures, higher luminosities and longer lifetimes compared to their corresponding non rotating ones, and since mass loss scales directly with the luminosity and inversely with the effective temperature, rotating models are expected to lose more mass than their respective non rotating counterparts. Moreover, mass loss in rotating models is enhanced due to the effect of the centrifugal force (see section \ref{franec}). Figure \ref{massHburn} shows that this expectation is fulfilled for masses $\rm M\leq 40~M_\odot$ and for the 120 \msun model while it is completely reversed for both the 60 \msun and the 80 \msun models. In order to understand such an "unexpected" result, it must be remembered that the mass loss rate provided by \cite{val00} has two threshold temperatures that mark dramatic changes in the mass loss rate. This means that if a star crosses or not one of these so called "bi-stability jumps", its mass loss rate may or may not change dramatically. Both the rotating 60 \msun and 80 \msun models evolve to higher effective temperatures compared to their non rotating counterparts and simply do not cross such threshold temperatures that mark the strong enhancement of the mass loss rate and therefore they do not lose as much mass as their respective non rotating models.

\begin{figure}
\epsscale{.99}
\plotone{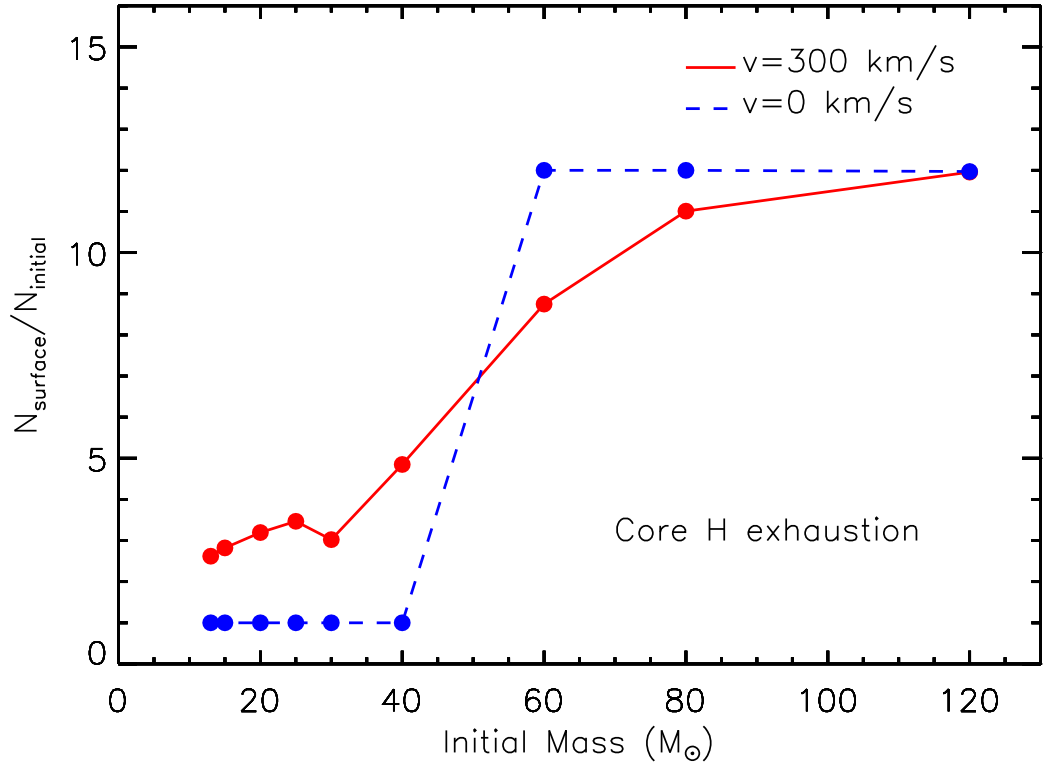}
\caption{Ratio between the final (at core H exhaustion) and the initial surface \nuk{N}{14} mass fraction 
as a function of the initial mass for rotating (red solid lines) and non rotating (blue dashed lines) models.}
\label{azotofinal}
\end{figure}

An important property of the models, worthwhile to be addressed, is the surface chemical composition at core H depletion. Figure \ref{azotofinal} shows the ratio between the final (at core H exhaustion) and the initial surface \nuk{N}{14} mass fraction as a function of the initial mass for rotating and non rotating models. 
No physical phenomenon can drive a change of the surface chemical composition during core H burning in classical non rotating models. In fact non rotating models with initial masses $\rm M\leq 40~M_\odot$ do not show any increase of the \nuk{N_{\rm surface}}{14}/\nuk{N_{\rm initial}}{14} ratio. Non rotating models with initial masses larger than 40 \msun, on the contrary, show a quite large \nuk{N_{\rm surface}}{14}/\nuk{N_{\rm initial}}{14} ratio, which is almost independent of the initial mass. The reason for such a \nuk{N}{14} enhancement is that mass loss becomes so efficient in these models that it peels off the outer envelope down to within the initial size of the convective core, in this way the products of the H burning may show up at the surface. Since the \nuk{N}{14} equilibrium abundance, resulting from the CNO processing, has a very mild dependence on the temperature, all these stars show roughly the same \nuk{N_{\rm surface}}{14}/\nuk{N_{\rm initial}}{14} value at the end of core H burning. 

Rotating models with initial masses lower than 40 \msun behave differently because in this case rotationally induced mixing is able to bring some products of the H burning (see Figures \ref{confstru1} and \ref{confstru2}) so that a sizable \nuk{N}{14} enhancement may be obtained in this case. As the initial mass increases, the mass loss increases as well and hence a stronger \nuk{N}{14} enhancement is expected. This is clearly visible in Figure \ref{azotofinal} although the \nuk{N}{14} enhancement is not as extreme as in the non rotating models. The reason is that stars in the mass interval 60 \msun to 80 \msun lose less mass than their non rotating counterparts (see Figure \ref{massHburn}) and therefore are not able to expose to the surface zones previously located within the convective core.

\begin{figure}
\epsscale{.99}
\plotone{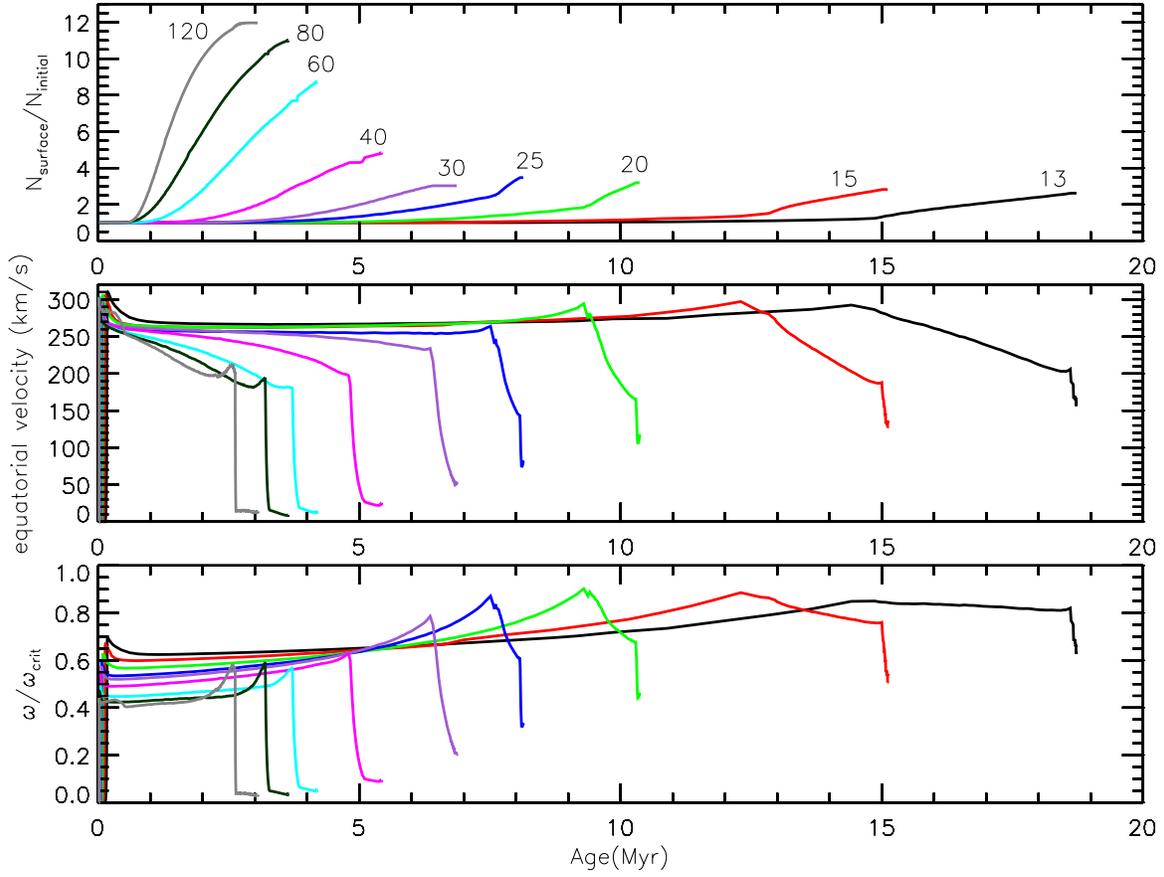}
\caption{Surface \nuk{N}{14} mass fraction normalized to the initial value (top panel), equatorial velocity $v$ (middle panel) and $\omega/\omega_{\rm crit}$ ratio (bottom panel) as a function of the H burning lifetime.}
\label{nitrotemp}
\end{figure}

Since Figure \ref{azotofinal} provides the surface chemical composition at core H depletion, it is not much useful to compare the expected surface properties of rotating models and the observed abundances in O-type and B-type stars. In fact the probability to see a star in a given evolutionary phase is proportional to the time spent in that phase. Therefore a much more meaningful information, for this purposes, is the temporal evolution of the surface \nuk{N}{14} mass fraction during core H burning. Figure \ref{nitrotemp} shows the surface \nuk{N}{14} mass fraction normalized to the initial value (top panel), the equatorial velocity $v$ (middle panel) and the $\omega/\omega_{crit}$ ratio (bottom panel) as a function of the H burning lifetime. This figure shows that the \nuk{N}{14} surface enrichment does not occur at the beginning but, on the contrary, in an advanced stage of core H burning. There is also a dependence on the initial mass. The smaller the mass the later the surface enrichment: in the 13-15 \msun models the enrichment occurs during the latest 20\% or so of core H burning. As the initial mass increases, the beginning of the surface enrichment extends over a progressively larger fraction of the H-burning lifetime. In the extreme case of the 120 \msun model, the surface enrichment starts just after the first 20\% of the core H-burning. The surface velocity remains roughly constant during most of the H burning, a small slowdown (20\% at most) occurring only in the more massive models. A comparison between the temporal evolution of the surface \nuk{N}{14} mass fraction and of the rotational velocity implies that in a steady state situation in which all stars rotate initially at 300 km/s, one should expect the following: (1) the majority of the stars (for any reasonable IMF, the lower mass stars are the most numerous ones) share a similar velocity, i.e. the initial one, and does not show any substantial surface \nuk{N}{14} enrichment; (2) the small fraction of stars showing a sizable \nuk{N}{14} enhancement should span a range of surface velocities between $\sim 250$ and $\sim 150$ km/s, because the surface \nuk{N}{14} enrichment occurs during the late stages of core H burning  which is characterized by the largest variation (decreasing) of the surface velocity.

Since the distribution of stars in the "(\nuk{N}{14}/\nuk{N_{\rm ini}}{14}$\rm)_{surf}$ -$\rm v~\sin(i)$" diagram (where $\rm v~\sin(i)$ is the projected rotational velocity, "i" being the angle between the rotational axis and the line of sight) plays a pivotal role in the comparison between the predictions provided by rotating models and the observed properties of the rotating stars, it is worth mentioning here that the temporal evolution of the surface abundance of \nuk{N}{14} depends on the basic scheme adopted to perform the rotationally induced mixing. To be clearer, in the previous section we showed in Figure \ref{checka04b} a plot of the run of the surface \nuk{N}{14} versus central H abundance for four different evolutionary tracks of a 20 \msun. The black line refers to our standard run (therefore computed within the advection-diffusion scheme discussed in section \ref{franec}) while the red, blue and green lines refer to three evolutions computed in the pure diffusion scheme but different choices for the free parameters $f_c$ and $f_\mu$ (see section \ref{calibration}). Note that, though all four tracks reach a similar final overabundance of \nuk{N}{14} at the end of core H burning, the change of the surface abundance with time (or, equivalently, the central H mass fraction) depends on the specific run. For example, while in the green run (pure diffusion with $f_c=0.2$ and $f_\mu=1.0$) the surface abundance begins to increase since almost the beginning of core H burning, in the red case (pure diffusion with $f_c=0.07$ and $f_\mu=0.0$) the surface abundance remains at its initial value for most of the time, raising at the required value just before the end of the H burning. Our standard track (black line) shows an intermediate behavior. Of course these differences will affect the expected distribution of stars in the (\nuk{N}{14}/\nuk{N_{\rm ini}}{14}$\rm)_{surf}$-$\rm V~\sin(i)$ diagram.

\begin{figure}
\epsscale{.89}
\plotone{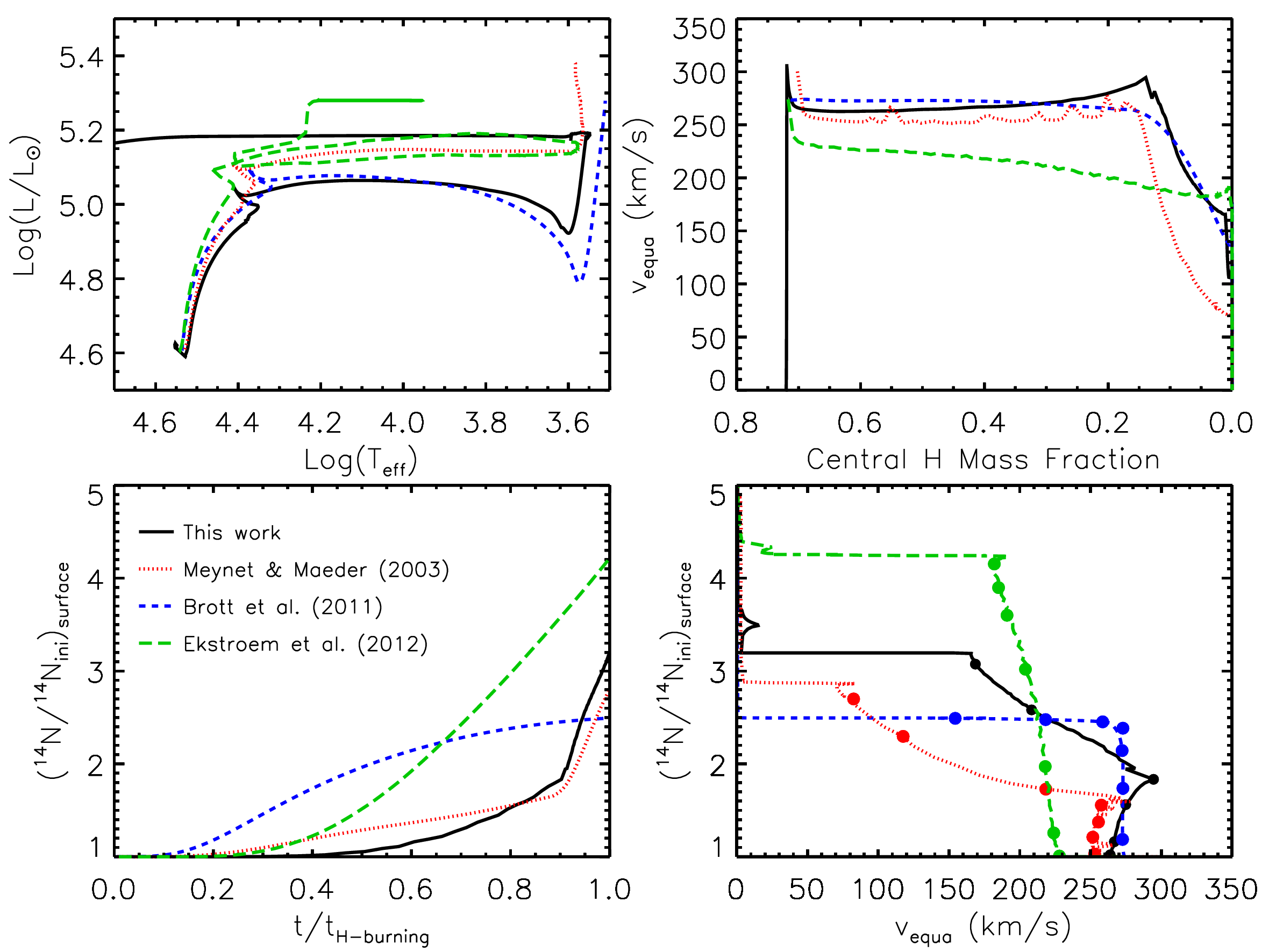}
\caption{Comparison among four 20 \msun models computed by different authors.
{\em Solid black line}: present work, Z=0.014, Y=0.265, $\rm v_{\rm ini}=300~{\rm km/s}$, $\rm X(C)_{\rm ini}=2.35~10^{-3}$ and $\rm X(N)_{\rm ini}=6.95~10^{-4}$.
{\em Dotted red line}: \cite{mm03}, Z=0.02, Y=0.275, $\rm v_{\rm ini}=300~{\rm km/s}$, $\rm X(C)_{\rm ini}=3.42~10^{-3}$ and $\rm X(N)_{\rm ini}=1.06~10^{-3}$.
{\em Dashed blue line}: \cite{brott11}, Z=0.008, Y=0.2638, $\rm v_{\rm ini}=273~{\rm km/s}$, $\rm X(C)_{\rm ini}=1.16~10^{-3}$ and $\rm X(N)_{\rm ini}=4.43~10^{-4}$.
{\em Long-Dashed green line}: \cite{ekstroemetal12}, Z=0.014, Y=0.266, $\rm v_{\rm ini}=270~{\rm km/s}$, $\rm X(C)_{\rm ini}=2.28~10^{-3}$ and $\rm X(N)_{\rm ini}=6.58~10^{-4}$.
The following properties are shown for each model: 
evolutionary path in the HR diagram (upper left panel); evolution of the equatorial velocity (upper right panel)
as a function of the central H mass fraction:
evolution of the surface \nuk{N}{14} mass fraction (normalized to the initial one) as a function of the
fraction of core H burning lifetime (lower left panel); evolution of the surface \nuk{N}{14} mass fraction (normalized to the initial one) as a function of the equatorial velocity (lower right panel).
}
\label{confazoto}
\end{figure}

Given the importance of such a diagram, we compare the evolutionary properties of our 20 \msun rotating model with those of similar models presently available in literature, namely, the one by \cite{mm03}, \cite{brott11} and \cite{ekstroemetal12} (see caption for more details on the models). The upper left panel of Figure \ref{confazoto} compares the path of these models in the HR diagram: the four models show a quite similar core H burning phase, all starting basically from the same location and diverging slightly during the evolution probably because of small differences in the extension of the convective core and in the efficiency of both the angular momentum transport and the rotationally induced mixing. Also the temporal evolution of the surface equatorial velocity is not much different (upper right panel, here the velocity is plotted as a function of the central H mass fraction), the largest discrepancy occurring in the \cite{ekstroemetal12} model: their model is the only one to show a systematic (even if not strong) drop of the surface velocity during core H burning. The temporal evolution of the surface N abundance (normalized to the initial one) is shown in the lower left panel. Both our model and the \cite{mm03} one show a very mild increase during most of the core H burning lifetime and a more consistent increase once the models enter the bi-stability jump. Viceversa, the \cite{brott11} model quickly reaches an asymptotic value that remains roughly constant during most of the core H burning lifetime. Note, however, that these models 
are significantly different from the others since they include the effects of magnetic fields on
the transport of the angular momentum (but not on the transport of the chemicals). The model taken from the \cite{ekstroemetal12} database shows an even different behavior, since it steeply increases all along the core H burning phase. The lower right panel shows the path of the four models in the (\nuk{N}{14}/\nuk{N_{\rm ini}}{14}$\rm)_{surf}$-$\rm V_{\rm eq}$ diagram with superimposed 7 dots along each track, marking the 20\%, 40\%, 60\%, 80\%, 90\%, 95\% and 100\% of the core H burning lifetime ($\rm t_H$), respectively. With the aid of these time-slices we can easily compare the predictions of the four different models. Our track (black line) predicts that a 20 \msun star remains $\sim 90\%$ of its $\rm t_H$ at basically its initial rotational velocity, with a surface N overabundance below a factor of two. Only the last $\sim 5\%$ of $\rm t_H$ is spent by this model with a relatively lower surface velocity (of the order of 150-200 km/s) and a higher N overabundance (between a factor of 2 and 3). The model of \cite{mm03} shows a similar behavior but with the difference that the typical rotational velocity during the last $\sim 5\%$ of its $\rm t_H$ is of the order of 50-100 km/s. The model provided by \cite{brott11} predicts a quite large spread of N overabundances at roughly constant equatorial velocity ($\sim 270~{\rm km/s}$, i.e., the initial one) for about 90\% of $\rm t_H$ and then a mild spread of the rotational velocities (150-250 km/s) at roughly constant N overabundance for the remaining $\sim 10\%$ of its $\rm t_H$. The behavior of the \cite{ekstroemetal12} model is even more extreme in the sense that it predicts a wide spread of N overabundances but basically at constant equatorial velocity all along the core H burning phase. These different behaviors of the four models will obviously reflect in a different distribution of the stars in the \nuk{N}{14}/\nuk{N_{\rm ini}}{14}$\rm)_{surf}$-$\rm V_{\rm eq}$ diagram. This issue will be discussed in a forthcoming paper.

\begin{figure}
\epsscale{.99}
\plotone{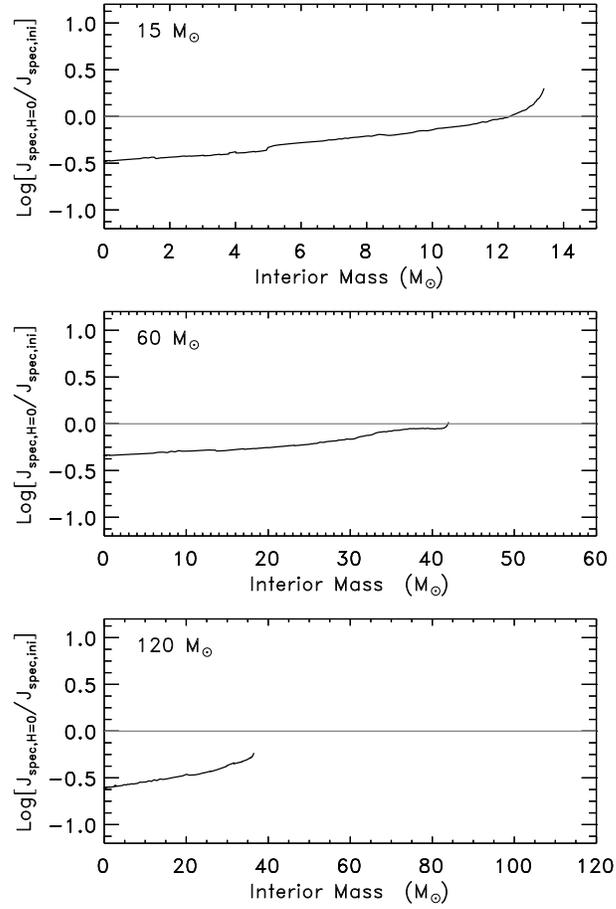}
\caption{Logarithm of the ratio between the final and initial specific angular momentum, as a function of the mass coordinate, at core H exhaustion, for three selected models, namely, 15 (upper panel), 60 (middle panel) and 120 \msun 
(lower panel).}
\label{deltamangH}
\end{figure}

Another property of rotating models which is worth discussing is the efficiency of the angular momentum transport during core H burning. Figure \ref{deltamangH} shows, for three selected models (i.e., 15, 60 and 120 \msun), the logarithm of the ratio between the final and initial specific angular momentum, as a function of the mass coordinate, at the end of core H burning. Inspection of this figure reveals that the angular momentum is transported from the innermost zones towards the outer ones and that it is eventually lost by stellar wind. As a result of these combined processes, rotating models lose roughly between $\sim 50\%$ and $\sim 75\%$ of their initial angular momentum during core H burning.

Before closing this section let us briefly mention that the more massive stars may lose such a large fraction of their initial mass during core H burning, that they become Wolf-Rayet (WR) \citep[see][for the definition of the various WR stages]{lc06} even before leaving the Main Sequence. The minimum mass to become a WR already during core H burning is 120 \msun, for a set of models with a Main Sequence equatorial rotational velocity of 300 km/s, while it drops to 80 \msun in the non rotating case. The rotating 120 \msun enters the WNL stage when the central H mass fraction is $\sim 0.15$ while the non rotating 80 and 120 \msun models enter the WNL stage at $\rm H_c \sim 0.01$ and $\rm H_c \sim 0.08$, respectively. It goes without saying that such a limiting mass would vary in the range 80-120\msun as the initial rotational velocity spans the range 0-300 km/s. This means that it is not straightforward to define unambiguously such (actually any - see below) a limiting mass in a realistic case in which a generation of stars is expected to show a flat (i.e. constant) Initial Distribution of ROtational Velocities (hereinafter IDROV). Note that in principle the IDROV is two-dimensional since the initial rotational velocity may depend either on the mass but also vary from star to star of the {\it same} mass. We'll come back to this point below. 

\section{Core He burning}

The first step that ferries a star from the end of core H burning to the central He ignition is the contraction and heating of the core, up to the typical He burning temperatures ($T\sim 200~{\rm MK}$), coupled to an expansion of the H-rich mantle. During this phase the star evolves from a blue supergiant (BSG) - high temperature - configuration, to a red supergiant (RSG) - low temperature - structure. The transition from a BSG to a RSG configuration is mainly controlled by the H-rich envelope. If the H-rich envelope is thick "enough", the star keeps a RSG structure while, on the contrary, as soon as the mass of the H-rich envelope decreases below a critical threshold value the star turns to a BSG configuration. By the way let us remind the reader that we are discussing solar metallicity stars; at lower metallicity a BSG structure may be retained even in presence of massive H-rich mantles. In the present set of models only the two (rotating and non rotating) 120 \msun fail to reach a RSG configuration (Figure \ref{hrtot}) because of the very thin H-rich envelope; all other models, rotating or not, become RSG at the very beginning of core He burning. The timescale of the transition from a BSG to a RSG configuration after core H depletion can not be robustly determined on the basis of first principles. The reason is that it depends on the efficiency of the mixing in the region of variable H left by the receding convective core during core H burning (Figure \ref{confstru2}) that in turn is still highly uncertain. In fact while the strict adoption of the Schwarzschild would imply that this region is unstable, the stabilizing effect of the $\mu$ gradient would keep these layers stable (Ledoux criterion). This zone is often referred to as the "semiconvective" region. If the Schwarzschild criterion for convection is adopted, these layers are mixed on a dynamical timescale and the redward excursion occurs on a nuclear timescale. This would imply that the region between the MS and the Red Giant Branch (RGB) in the HR diagram should be well populated. On the contrary, the adoption of the Ledoux criterion prevents a mixing on a dynamical timescale and in this case the redward evolution occurs on the much faster Kelvin-Helmoltz timescale. In this case one would expect very few stars (basically a gap) in the HR diagram between the MS and the RGB. Unfortunately, the Ledoux criterion is not that robust because the gradient of mean molecular weight ($\nabla_\mu$), that stabilizes the zone against convection, could be (at least at small scales) destroyed by any kind of stochastic turbulence, favoring the dynamical mixing that, in turn, would reduce even more $\nabla_\mu$. This is clearly an unstable situation. The physics of this phenomenon has been studied by many authors by means of linear-stability analyses. For example \cite{kato66} showed that infinitesimal perturbations would grow in this region on a thermal diffusion timescale, due to heat dissipation processes, with the net result of mixing eventually the whole region. Unfortunately, the timescale over which these perturbations become finite as well as the real time-scale of the mixing process is still unknown. Hence the only guidance we have to treat these layers comes from the observations which clearly show a shortage in the number of stars between the MS and the RGB \citep[][and references therein]{massey2003}. This implies that at most a small efficiency of the mixing is allowed by the observations. In the present models we treat the semiconvective region following the same formalism introduced by \cite{langer91} and we fix the free parameter $\alpha_{\rm semi}$ to the largest value that allows all the models to become RSG at the very beginning of core He burning: in our models it amounts to $\alpha_{\rm semi}=0.02$.

\begin{figure}
\epsscale{.99}
\plotone{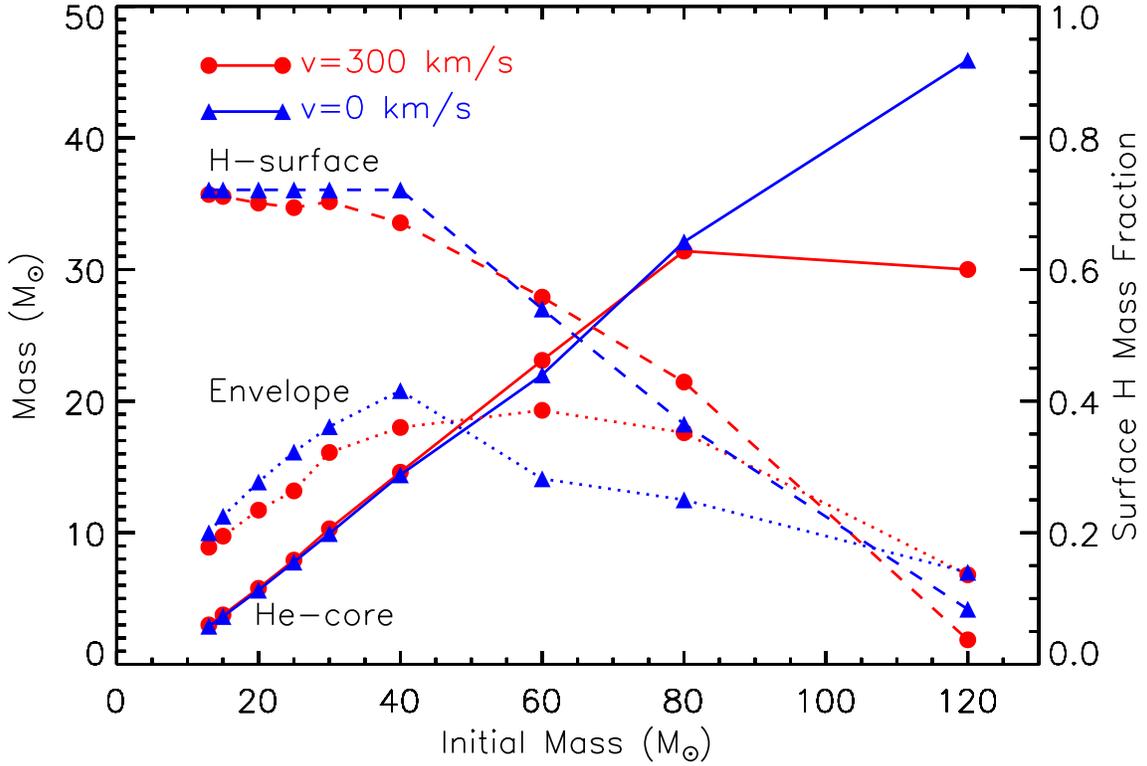}
\caption{Selected properties of rotating (red filled circles) and non rotating (blue filled triangles)
models at core H exhaustion: the He core mass (solid lines), the mass of the envelope (dotted lines) and the surface H mass fraction
(dahsed lines, reported on the right y-axis).}
\label{mhecore}
\end{figure}

\begin{figure}
\epsscale{.99}
\plotone{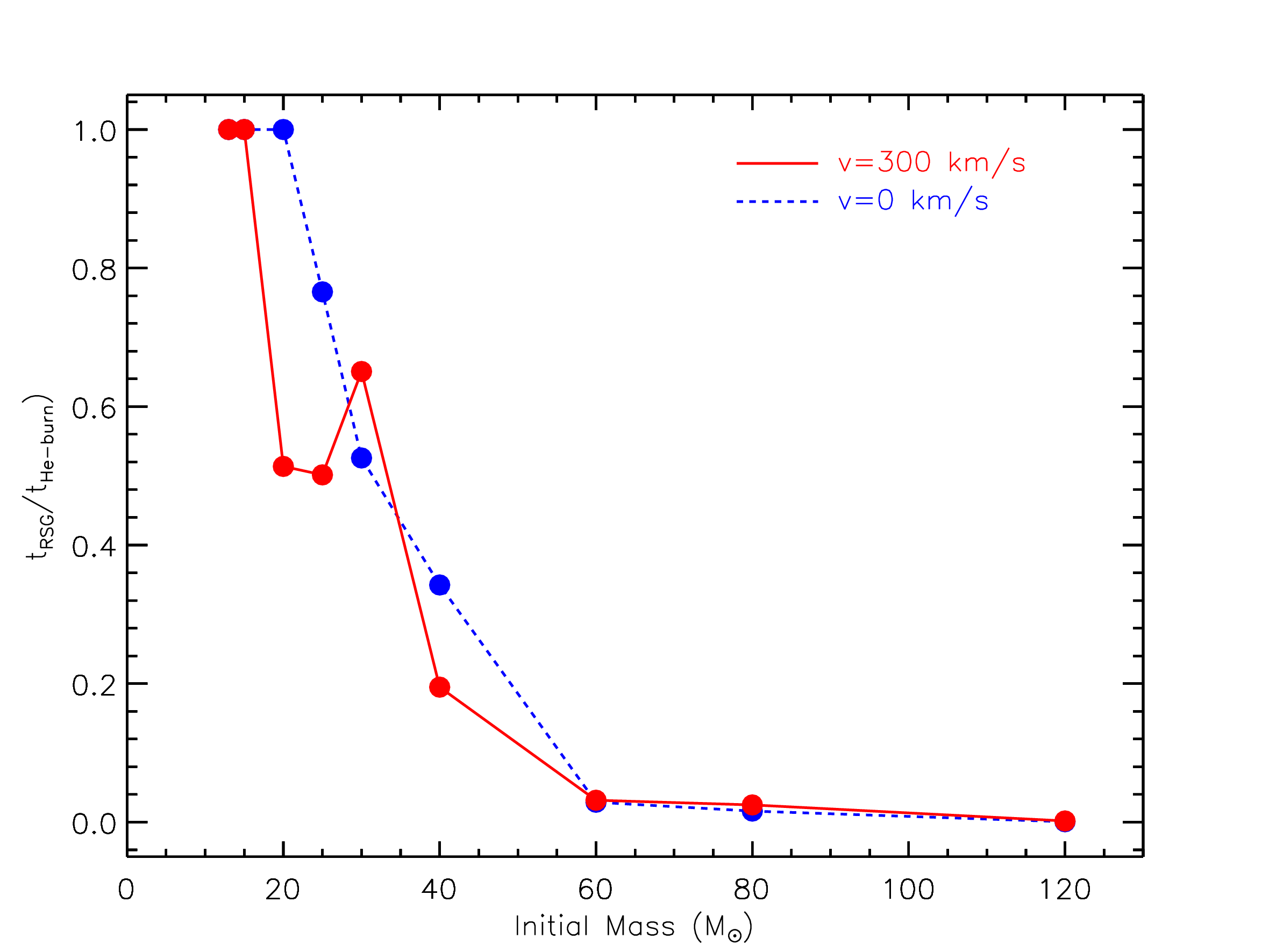}
\caption{Fractional time spent by each model as RSG with respect to its total core He burning lifetime. Solid lines refer to
rotating models while dashed lines refer to non rotating models.}
\label{red_blue}
\end{figure}

Figure \ref{mhecore} summarizes some selected properties of the models at core H exhaustion and shows that the He core masses are almost independent of rotation, the only exceptions being the 120 \msun since the rotating star lose much more mass during the MS than its non rotating counterpart (see previous section) entering the Wolf-Rayet phase when the central H abundance is still relatively high: the very efficient mass loss associated to the WR phase has therefore much more time to operate. As already discussed above, the position of a model in the HR diagram in core He burning depends on the mass size of the H-rich envelope (defined as the amount of mass above the He core mass). Figure \ref{mhecore} (dotted lines) also shows that rotating and not rotating models reach the end of the core H burning with substantially different H-rich mantles. The consequence of this difference on the evolutionary path of the models in the HR diagram during core He burning is shown in Figure \ref{red_blue}. In particular, all models (rotating and not) with $M\le 15~{\rm M_\odot}$ spend all their core He burning as RSGs since in both cases the mass of the mantle remains above the critical value needed for a blueward evolution. On the contrary, all models with mass $M\ge 60~{\rm M_\odot}$ evolve to the blue at the very beginning of the He burning, regardless of the mass size of the envelope, because of the huge mass loss typical of this mass interval that is able to eject all the mantle anyway. The largest differences between rotating and non rotating models manifest in the mass interval between the 20 and the 40 \msun. The reason is that stars in this mass interval spend part of their central He burning lifetime as RSG and part as BSG. As already discussed above the transition occurs once the mass size of the envelope has greatly reduced and since rotating models reach the He ignition with smaller envelope masses, they will lose their H rich envelope earlier than their non rotating counterparts (that have more massive H-rich envelopes). Hence the ratio between the time spent as RGB and the total time spent in central He burning is systematically smaller in presence of rotation. Also the minimum mass that experiences a blueward evolution (which means the smallest mass that becomes WR, see below) reduces in presence of rotation. Figure \ref{red_blue} shows the $\rm t_{RSG}/t_{He-burn}$ ratio as a function of the initial mass for both sets of models. In the specific case of stars rotating initially at 300 km/s the minimum mass that becomes WR is smaller than 20 \msun (to be compared to a limiting mass of the order of 25 \msun obtained in the non rotating case). Of course since not all stars are born with the same rotational velocity, the concept of a {\it unique} limiting mass loses of meaning in presence of rotation and should be replaced by the idea that a {\it spread} of limiting masses exists due to the variety of possible initial rotational velocities. Let us eventually note that non rotating models in the interval 20-35 \msun experience a blue loop during core He burning, turning back towards a RSG/YSG (YSG stands for Yellow Supergiant) configuration at core He depletion. 

\begin{figure}
\epsscale{.99}
\plotone{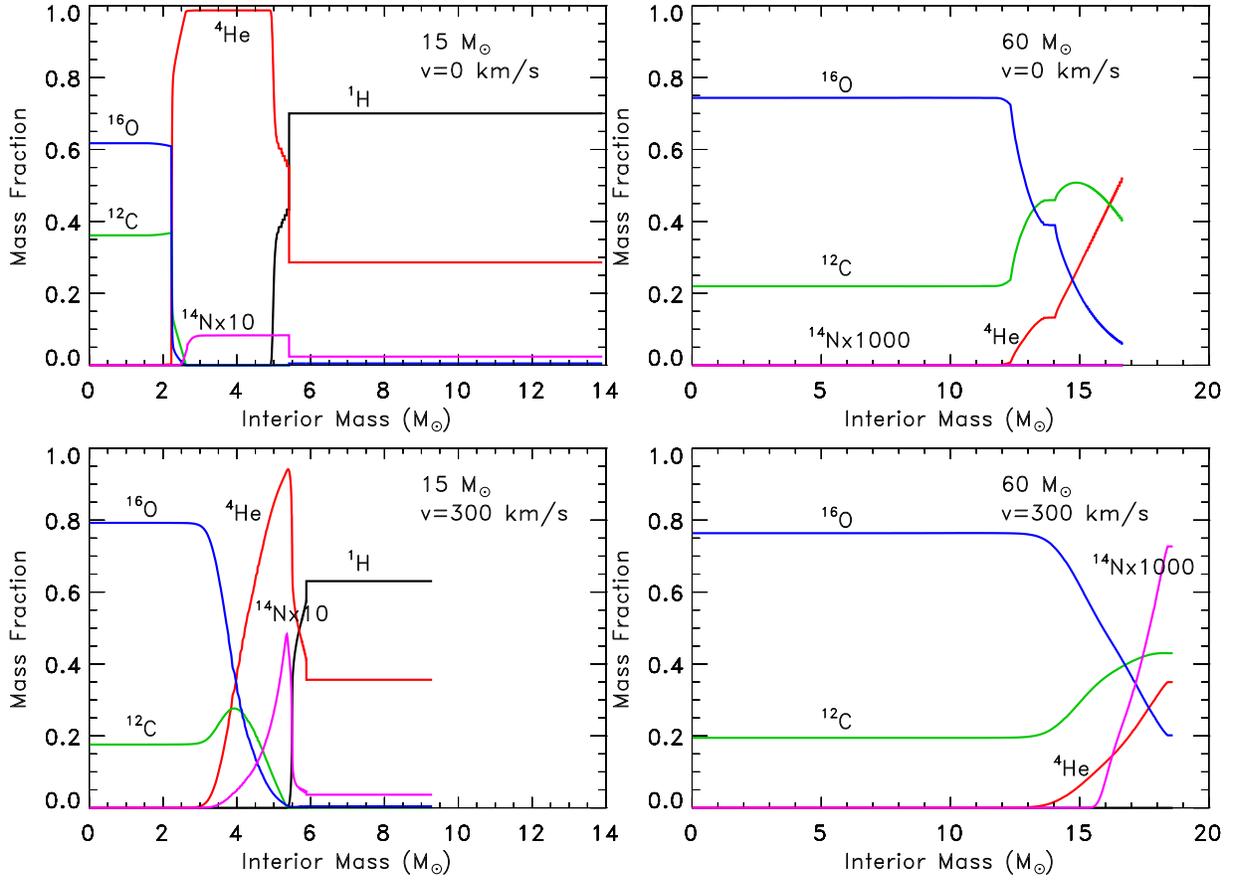}
\caption{Internal chemical composition for four selected models at core He exhaustion: non rotating (upper left panel) and rotating (lower left panel)
15 \msun models; non rotating (upper right panel) and rotating (lower right panel) 60 \msun models.}
\label{1560endHe}
\end{figure}

Let us now turn to the internal properties of these models starting with the non rotating models with mass $M\leq40~{\rm M_\odot}$. These models are characterized by a He convective core that either advances or remains stable in mass. If the Schwarzschild criterion were adopted, the internal chemical composition of these stars, at core He depletion, would be characterized by a sharp chemical discontinuity marking the maximum extension of the He convective core. Behind such a discontinuity, the chemical composition would be
homogeneous and dominated by \nuk{C}{12} and \nuk{O}{16}, i.e., the main ashes of He burning. Outside the chemical discontinuity, the chemical composition would be dominated by \nuk{He}{4} and \nuk{N}{14}, i.e. the main products of core H burning, and would be homogeneous as a result of the convective mixing during core H burning. The adoption of the Ledoux criterium, viceversa, creates a small region of variable chemical composition at the border of the convective core in which \nuk{C}{12} is already enhanced with respect to the pre He burning abundance but the \nuk{N}{14} has not yet been destroyed by $\alpha$ captures. This special region remains untouched up to the end of the central He burning and is of great importance because its chemical composition resembles the one observed on the surface of a quite rare subset of WR stars, namely the WNC stars  ($0.1<{\rm (C/N)_{surf}}<10$) (upper left panel of Figure \ref{1560endHe}). In these models, however, mass loss is not strong enough to expose these layers to the surface, hence none of them will show up as WNC stars.

Non rotating stars with mass larger than 40 \msun, on the contrary, experience a strong mass loss after core H burning so that they completely lose the H rich envelope and a fraction of the He core during core He burning. Such an occurrence deeply affects the evolution of these stars. The upper right panel of Figure \ref{1560endHe} shows the internal chemical composition of a non rotating 60 \msun star, taken as representative of the high mass models. The He profile looks quite different from the one shown by the 15 \msun. In this case, in fact, the continuous shrinking of the He convective core caused by the progressive reduction of the He core mass induced by the strong mass loss (note that a similar effect could be obtained in the case of a Roche lobe overflow) leads to the formation of a shallow He profile. By the way, this is a clear demonstration of the influence of the He core mass size on the evolution of the internal regions: the smaller the (He core) mass the smaller the size of the convective core.

Let's now turn to the rotating models (lower panels in Figure \ref{1560endHe}). In the 15 \msun case, the He profile shows now a pronounced slope within the whole He core. This is not the consequence of a reduction of the He core (that is not eroded by mass loss in this case), but the result of the work done by the rotationally induced mixing that partially spreads the products of the He burning within the He core and also slightly within the tail of the H burning shell. This slow ingestion of fuel (He in this case) within the convective core has the visible consequence of lowering significantly the $\rm ^{12}C/^{16}O$ ratio and of increasing the mass of the CO core at the end of core He burning \citep[see][for a detailed discussion of the influence of the mixing on the $\rm ^{12}C/^{16}O$ ratio at core He exhaustion]{imb01},  while the contemporaneous penetration of fresh $\rm ^{12}C$ into the tail of the H burning shell activates the production of a locally consistent amount of primary $\rm ^{14}N$. Induced rotational mixing operates also in more massive models at the beginning of core He burning and produces the same effects discussed for the 15 \msun, but then mass loss begins to reduce progressively the He core so that the chemical composition present within the He core at core He exhaustion results from the combined effect of the rotationally induced mixing (operating during the early stages of core He burning) and the progressive reduction of the He core due to mass loss (working at later stages of core He burning).

Tables 3 and 4 show the time spent by all the present models in each specific WR stage. The various columns show: 1) the initial mass, 2) the time spent by the model as O-type star (in yr), 3) the time spent globally as WR star (in yr), 4) the time spent as WNL (in yr), 5) the central abundance, between brackets, when the star enters the WNL phase, 6) the time spent as WNE (in yr), 7) the central abundance, between brackets, when the star enters the WNE phase, 8) the time spent as WNC (in yr), 9) the central abundance, between brackets, when the star enters the WNC phase, 10) the time spent as WCO (in yr) and 11) the central abundance, between brackets, when the star enters the WCO phase. Inspection of these tables shows that the minimum mass for the formation of a WR star is of the order of $15-20~{\rm M_\odot}$ if $v_{ini}=300~{\rm km/s}$, value that raises to $20-25~{\rm M_\odot}$ in the non rotating case. It is interesting to note how strongly the WNE and the WNC stages are influenced by rotation when the initial rotational velocity is $v_{ini}=300~{\rm km/s}$. While only a minor fraction of the region confined between the convective core and the border of the He core is enriched in $\rm ^{12}C$ in the non rotating case (see above), the rotationally induced mixing spreads out fresh $\rm ^{12}C$ produced in the core within the whole He core (raising the C/N ratio above 0.1) so that no He rich layer resembles any more the typical composition of a WNE star but that of a WNC one (Figure \ref{comp60He}). More specifically, all rotating models with mass $M<120~M_\odot$ change directly from a WNL to a WNC stage (with the only exception of the $20~M_\odot$) and fully skip the WNE stage. On the contrary, both the WNE and WNC stages are populated in absence of rotation, being 40 \msun and 60 \msun the minimum masses that become WNE and WNC respectively.

\begin{figure}
\epsscale{.99}
\plotone{fig16.eps}
\caption{Internal chemical composition prior to the entrance in the WR stage for the rotating 60 \msun (left panel) and for the non rotating 60 \msun. The diffusion coefficients are also shown in the figure and reported in the secondary axis.}
\label{comp60He}
\end{figure}


\begin{figure}
\epsscale{.99}
\plotone{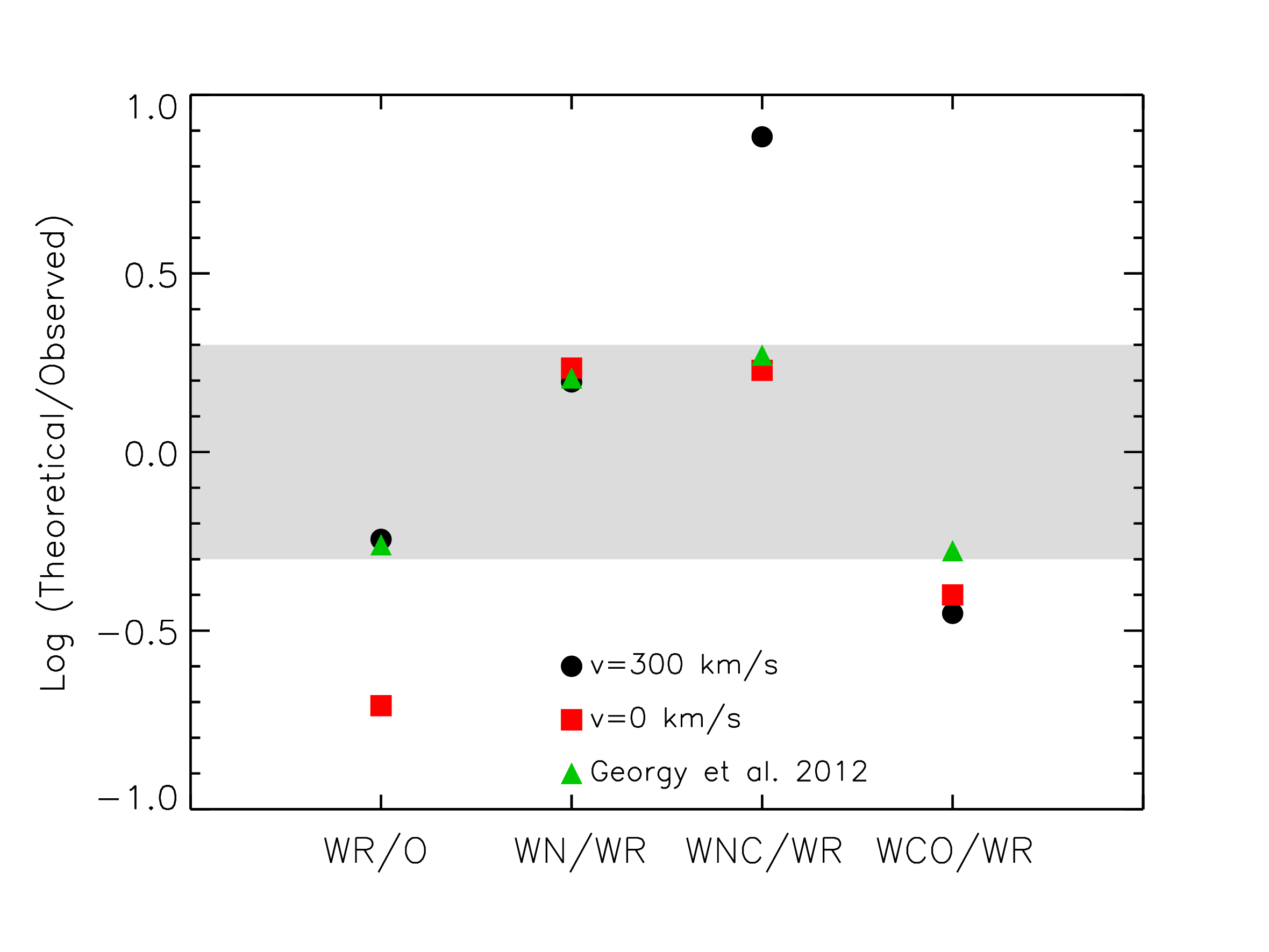}
\caption{Comparison between the theoretical WR/O and WR-subtypes/WR ratios and the observed ones for the solar neighborhood. The observed WR/WR-subtypes ratios have been taken from the compilation of \cite{georgy12}. The grey band  marks a region corresponding to an uncertainty of a factor of 2. The filled black circles and the filled red squares refer to the present rotating and non rotating models, respectively. The filled green triangles refer to the rotating models provided by \cite{georgy12}.
}
\label{wrconf}
\end{figure}

Figure \ref{wrconf} shows a comparison between the expected and the observed numbers of Wolf-Rayet stars with respect to the O-type stars (WR/O) and the WR-subtypes/WR ratios for the two sets of models. By the way let us remind that this comparison is obtained by making the nonphysical assumption that the IDROV is flat and that the rotational velocity is 300 km/s. Note that the assumed IDROV is bidimentionally flat in the sense that we assume either that the initial surface equatorial velocity is independent on the mass but also that all stars of the same mass born with the same initial velocity (there is no evident reason for this). For this reason a more realistic prediction of the number ratios among the various WR-subtypes and O-type stars should be based on synthetic stellar population synthesis in which an IDROV as anchored as possible to the real world is taken into account.

Keeping in mind this general warning, Figure \ref{wrconf} shows a comparison between the expected ratios under the assumption of a steady state distribution of stars of solar metallicity and a Salpeter IMF ($n(m)=k\cdot m^{-\alpha}$) with $\alpha=2.35$. The observed WR/WR-subtypes ratios have been taken from the compilation of \cite{georgy12}

The grey band in Figure \ref{wrconf} marks a region corresponding to an uncertainty of a factor of 2. The integrated WN/WR and WCO/WR ratios are basically unaffected by rotation while both the WR/O and the WNC/WR ratios predicted by the present rotating models are a factor of 3 to 4 higher than in the non rotating case. A comparison between our theoretical predictions and those obtained by \cite{georgy12} shows an excellent agreement for the WR/O, WN/WR, and WCO/WR ratios (Figure \ref{wrconf}). Viceversa, the WNC/WR ratio differs by a factor of $\sim3-4$. Such a difference should come out from a different efficiency of the rotational mixing in the region between the He convective core and the He core, and from the presence and extension of the semiconvective region at the border of the He convective core (see above). The comparison between the observed and the predicted ratios is quite satisfactory in both the rotating and non rotating cases, at least within a factor of 2. The only significant discrepancies occur for the WR/O ratio, in the non rotating case, and for the WNC/WR ratio, in the rotating case. The large underestimated WR/O ratio, obtained in the non rotating models, means that the minimum mass for WR star formation is too large, a reduction of this limiting mass should significantly improve the fit. As far as the rotating models are concerned, the large overestimated WNC/WR ratio is connected to the very efficient rotationally induced mixing during core He burning that, as discussed above, produces an extended zone between the outer edge of the He convective core and the tail of the H burning shell, where both $\rm ^{12}C$ and $\rm ^{14}N$  are quite abundant. This efficient mixing is also responsible for the lack of WNE stars predicted in this case. As a final comment, it is interesting to note that, in spite of previous claims that rotation is necessary in order to predict a non zero population of WNC stars \citep{mm03}, our non rotating models predict a number of WNC stars that is compatible with the observed number.

\begin{figure}
\epsscale{.99}
\plotone{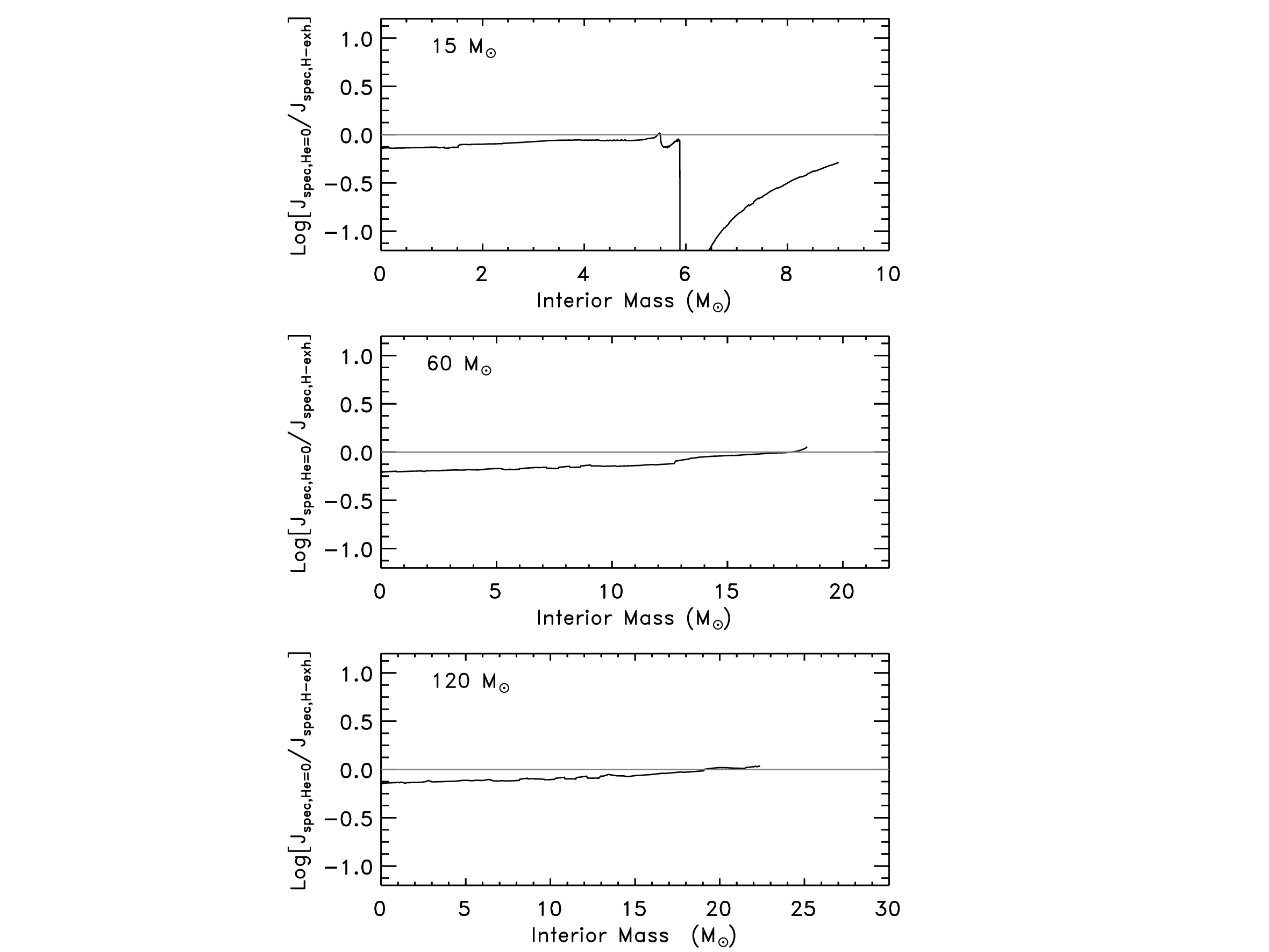}
\caption{Logarithm of the ratio between the final (at core He exhaustion) and initial (at core H exhaustion) specific angular momentum, as a function of the mass coordinate, for three selected models, namely, 15 (upper panel), 60 (middle panel) and 120 \msun 
(lower panel).}
\label{deltamangHe}
\end{figure}

Let us now turn to the evolution of the angular momentum and of the equatorial surface velocity during core He burning. Figure \ref{deltamangHe} shows the logarithm of the ratio between the final (at core He exhaustion) and the initial
(at core H exhaustion) specific angular momentum for three selected models, namely, 15 (upper panel), 60 (middle panel) and 120 \msun (lower panel). Inspection of Figure \ref{deltamangHe} reveals that these models lose a consistent fraction of their angular momentum also in He burning; models of lower mass lose roughly 10-15\% of the angular momentum present in the core, while the more massive models roughly halves the amount of angular momentum located in the interior. Models that burn He as RSGs (13 and 15 \msun) maintain a very small rotational velocity because of the very extended radii. Viceversa, those becoming BSGs experience a substantial increase of the rotational velocity due to the contraction and the consequent speed up of the outer layers. However, all these models converge towards similar structures at core He exhaustion, the final surface velocity being of the order of 100 km/s. The only exception is the 120 \msun: this model experiences such a huge mass loss (and angular momentum) during core H burning that the surface equatorial velocity at core He exhaustion drops to roughly 50 km/s.

After the core He depletion, He burning shifts in a shell and eventually a convective zone above the He burning shell forms. Such a He convective shell develops either in a region of variable He (i.e. in all rotating models and in the non rotating ones with $M\geq~40~{\rm M_\odot}$), or in a region where He is flat (i.e. the non rotating models with $M < ~40~{\rm M_\odot}$). In the first case the He convective shell turns out to be hotter then in the second one \citep{lc06}; such a different behavior will have profound and interesting implications on the production of some specific isotopes (see below). Before closing this section let us eventually point out that the He convective shell destroys most of the primary $\rm ^{14}N$ produced by the induced rotational mixing so that no primary N is ejected in the interstellar medium.

\section{The advanced burning phases}

The physical parameters driving the evolutionary properties of a massive star after the core He exhaustion are (1) the CO core mass (that takes the role of the total mass), (2) the $\rm ^{12}C$ mass fraction profile left by core He burning and, if the star rotates, (3) both the total and the interior distribution of the angular momentum. 

\begin{figure}
\epsscale{.99}
\plotone{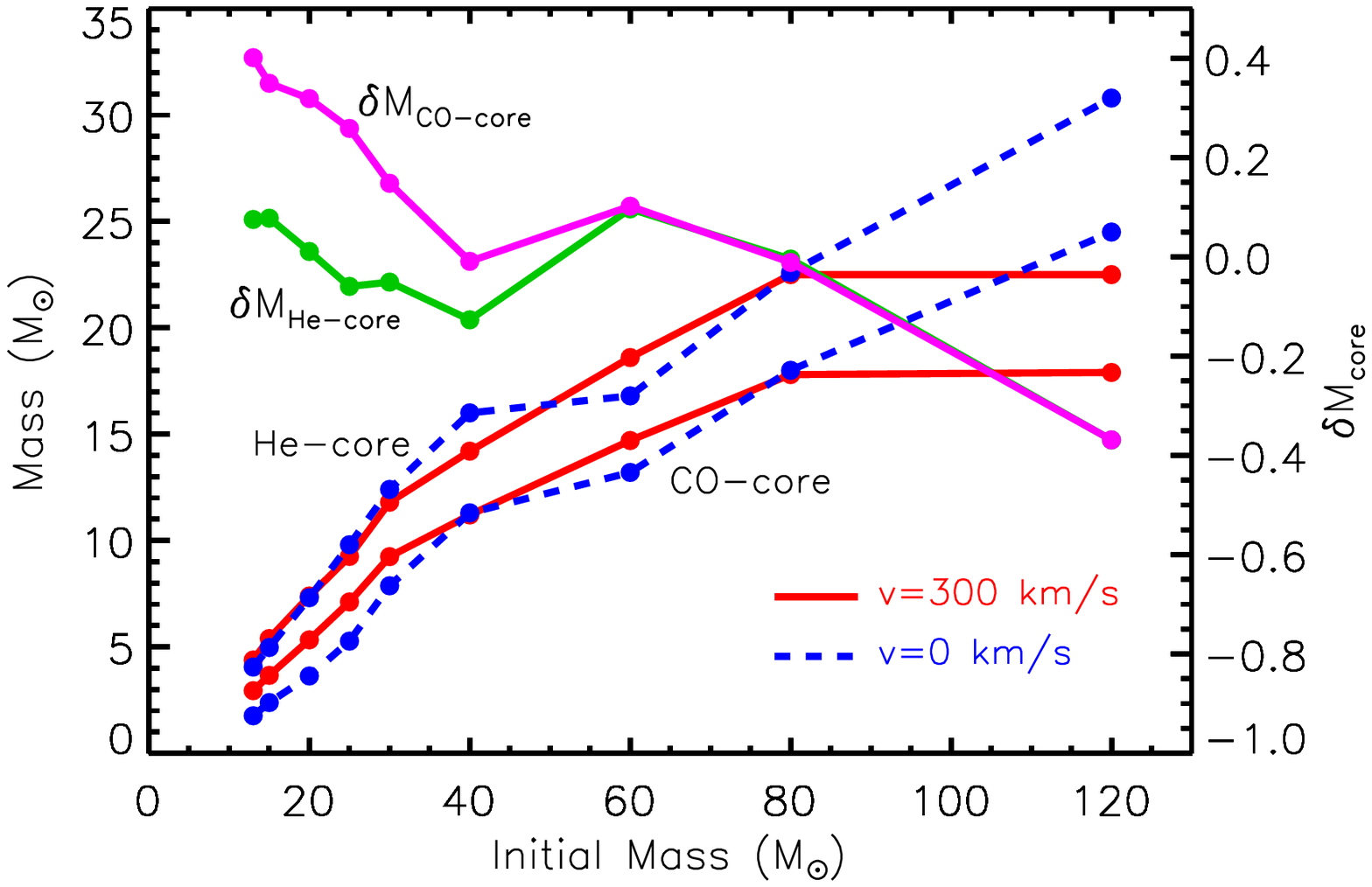}
\caption{
He core and CO core masses as a function of the initial mass for both the rotating (red solid lines) and the non rotating (blue dashed lines) models. Also shown are the relative differences, $\rm \delta M_{core}=(M_{core}^{rot}-M_{core}^{norot})/M_{core}^{rot}$, of both the He (green line) and the CO (magenta line) cores between rotating and non rotating of models.
}
\label{mcohe0}
\end{figure}

\begin{figure}
\epsscale{1.1}
\plottwo{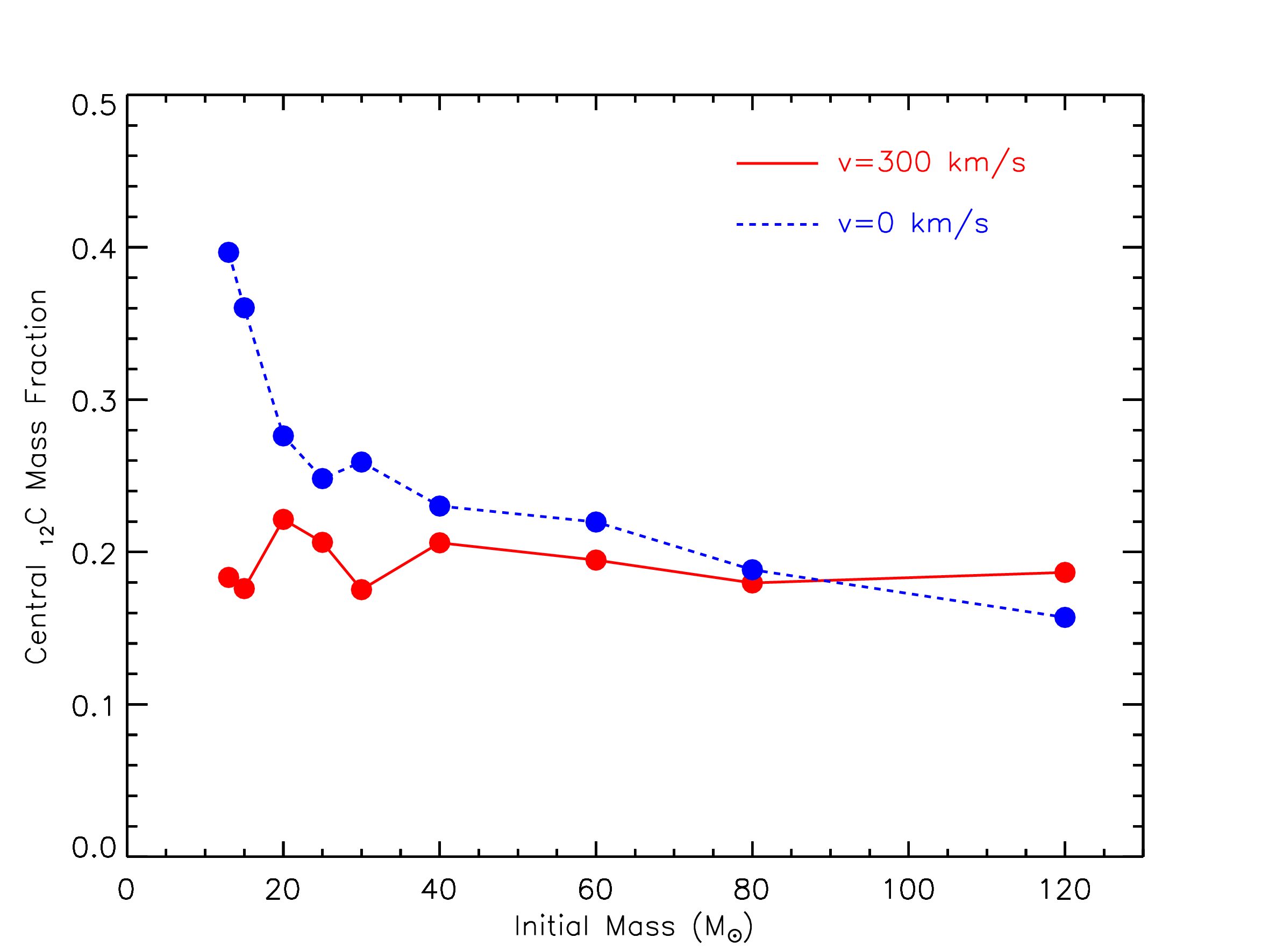}{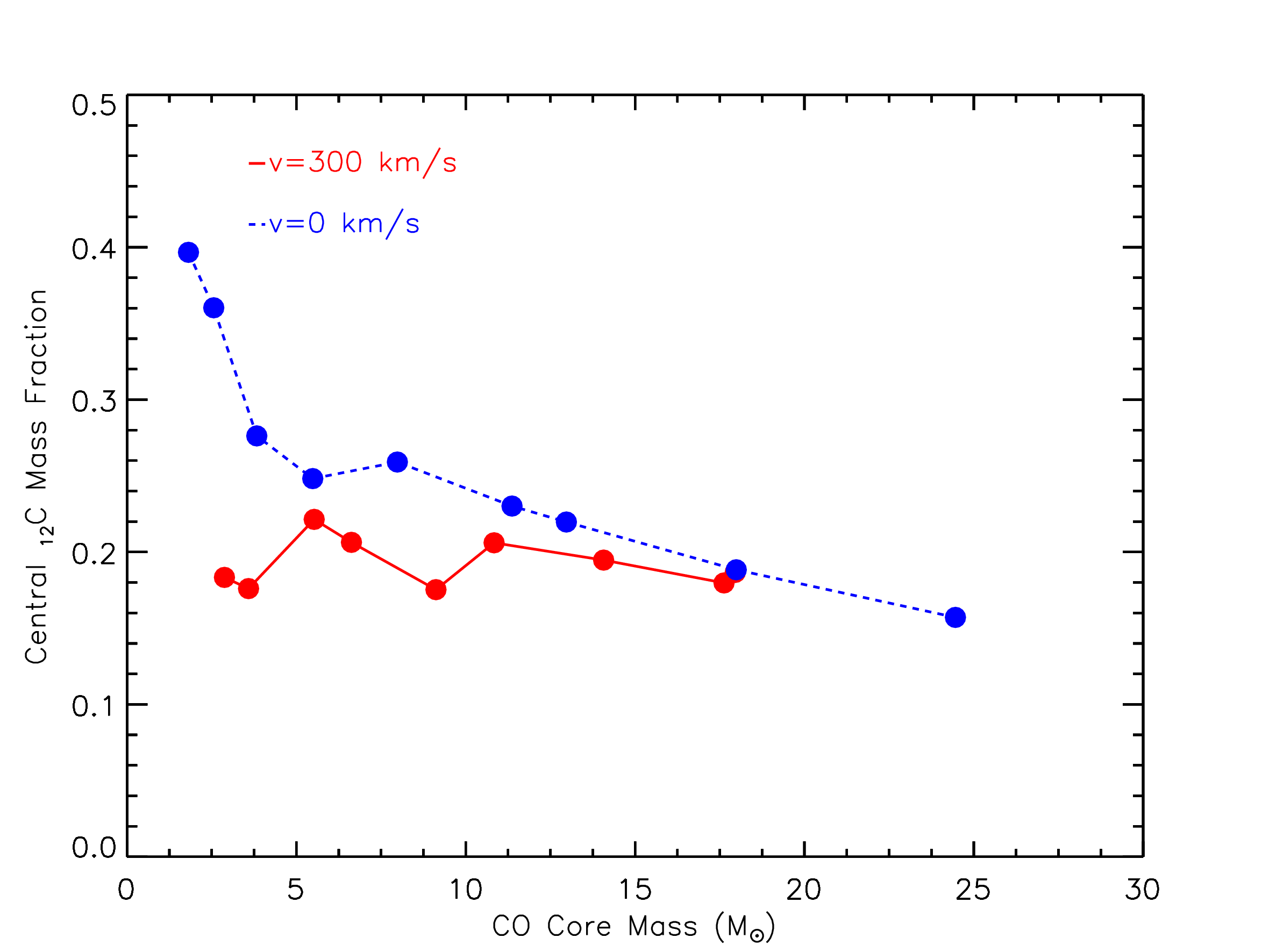}
\caption{Central $\rm ^{12}C$ mass fraction as a function of either the initial mass (left panel) and the CO core mass (right panel) for rotating (red solid lines) and non rotating (blue dashed lines) models.}
\label{ccen}
\end{figure}

\begin{figure}
\epsscale{.99}
\plotone{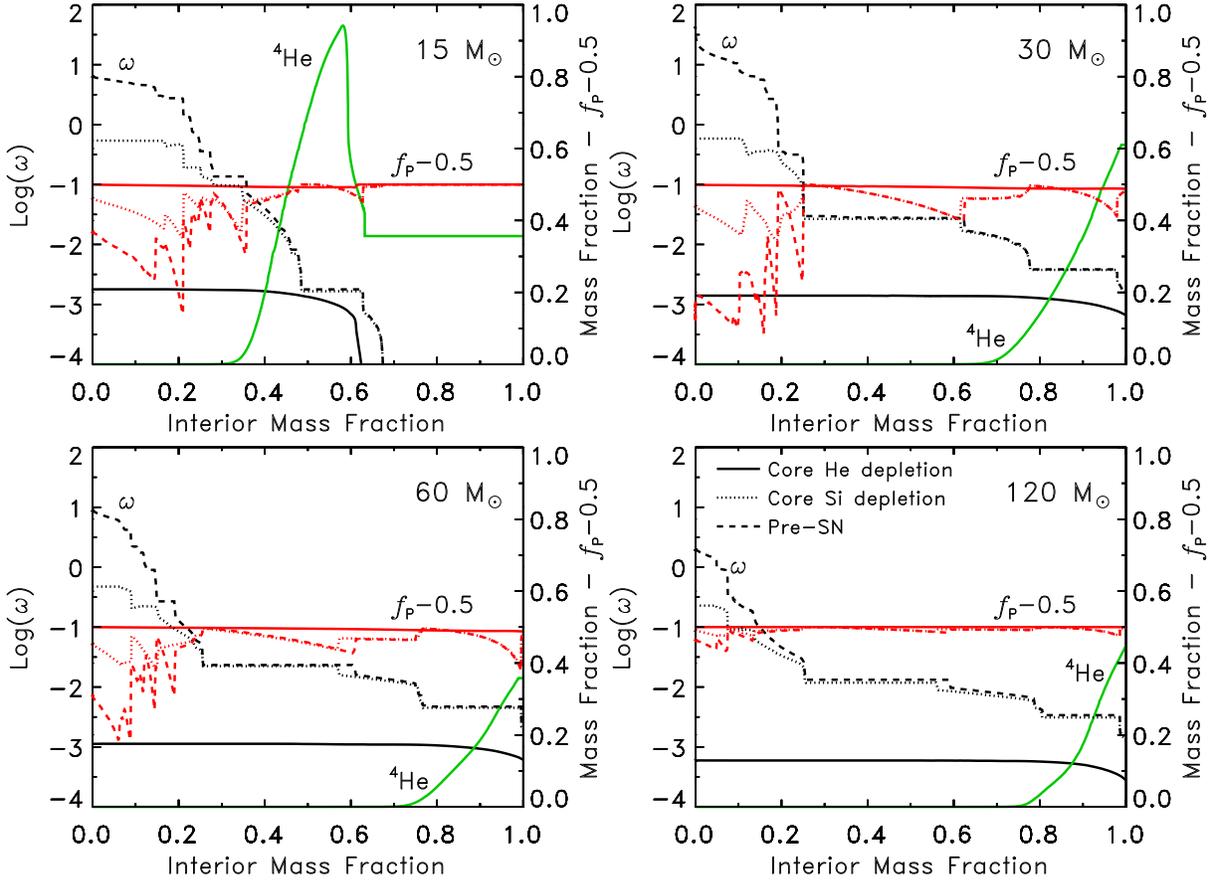}
\caption{Angular velocity (black lines) and "form factor" $f_P$ (red lines) as a function of the interior mass fraction at core He exhaustion (solid lines), core Si exhaustion (dotted lines) and presupernova stage (dashed lines), for four selected models, namely, 15 \msun (upper left panel), 30 \msun (upper right), 60 \msun (lower left panel) and 120 \msun (lower right panel). Also shown is the He mass fraction as a function of the interior mass at core He exhaustion (green lines).}
\label{omegahe0}
\end{figure}

\begin{figure}
\epsscale{.99}
\plotone{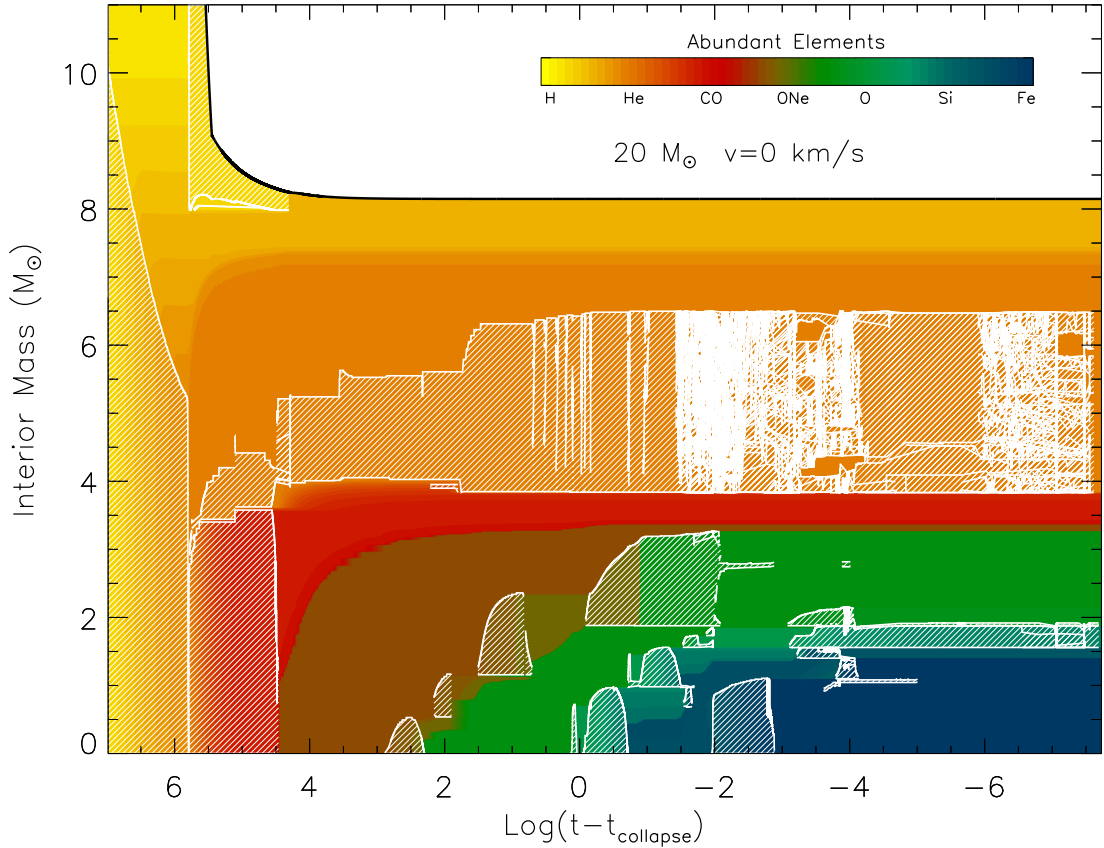}
\caption{Convective and composition history of a rotating 20 \msun model. Convective zones are marked by shaded areas while the chemical composition is coded as in the upper right color bar. The quantity on the x-axis is the logarithm of the residual time to the collapse in years while the quantity reported on the y-axis is the interior mass coordinate in solar masses.}
\label{kip020R}
\end{figure}

\begin{figure}
\epsscale{.99}
\plotone{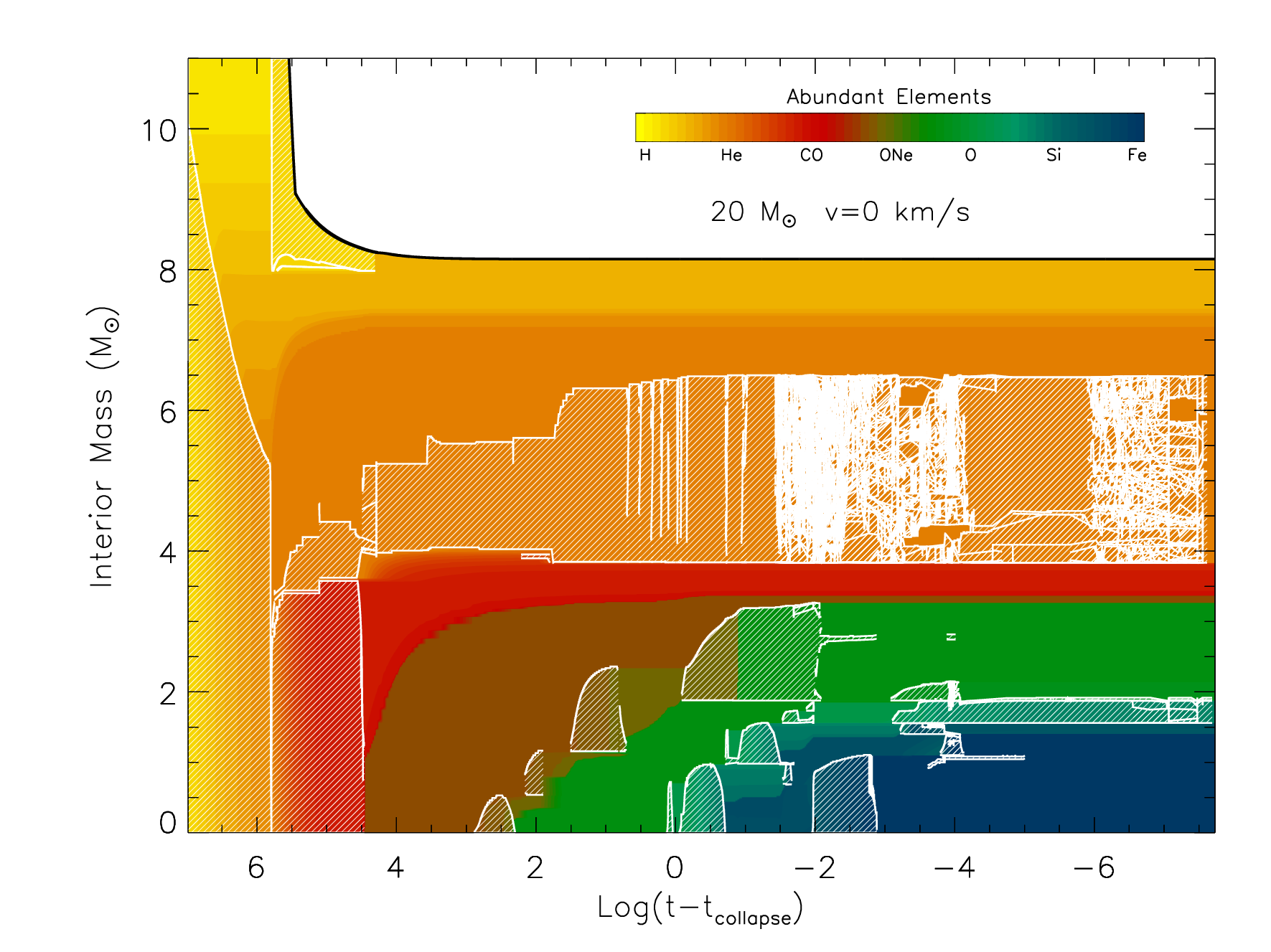}
\caption{Same as Figure \ref{kip020R} but for a non rotating 20 \msun model.}
\label{kip020N}
\end{figure}

These basic properties of the models at core He exhaustion, as a function of the initial mass, are shown in Figures \ref{mcohe0}, \ref{ccen} and \ref{omegahe0}. We have already shown in the previous section (Figure \ref{1560endHe}) that an important consequence of the rotationally induced mixing during core He burning is a lower $\rm ^{12}C$ abundance and a larger CO core mass at core He exhaustion, compared to the non rotating case (this is also readily evident from Figures \ref{mcohe0}, \ref{ccen}). These "effects", however, do not affect all the models in the same way but, on the contrary, they reach their maximum for the lowest masses and reduce progressively as the initial mass increases. The reason is partly due to the fact that we adopted an initial equatorial surface velocity equal for all the models, independent on the initial mass. Since this choice implies that also the initial $\omega/\omega_{\rm crit}$ reduces as the initial mass increases, the effects of rotation become progressively less important as well. Moreover, it must also be reminded that the evolutionary timescales decrease with the initial mass, hence the larger the mass the faster the evolution and hence the smaller the timescale over which the rotationally induced secular instabilities may operate. The black lines in Figure \ref{omegahe0} show the trend of the angular velocity as a function of the interior mass, for four selected rotating models at core He exhaustion. The basically flat profile shown by all the models within their CO core descends from the fact that the CO core basically corresponds to the He convective core and that we impose a solid body rotation in the convective regions. Note the sharp reduction of the angular velocity in the envelope of the 15 \msun model which is the only one, among the four, that is a RSG at this stage.

\begin{figure}
\epsscale{.99}
\plotone{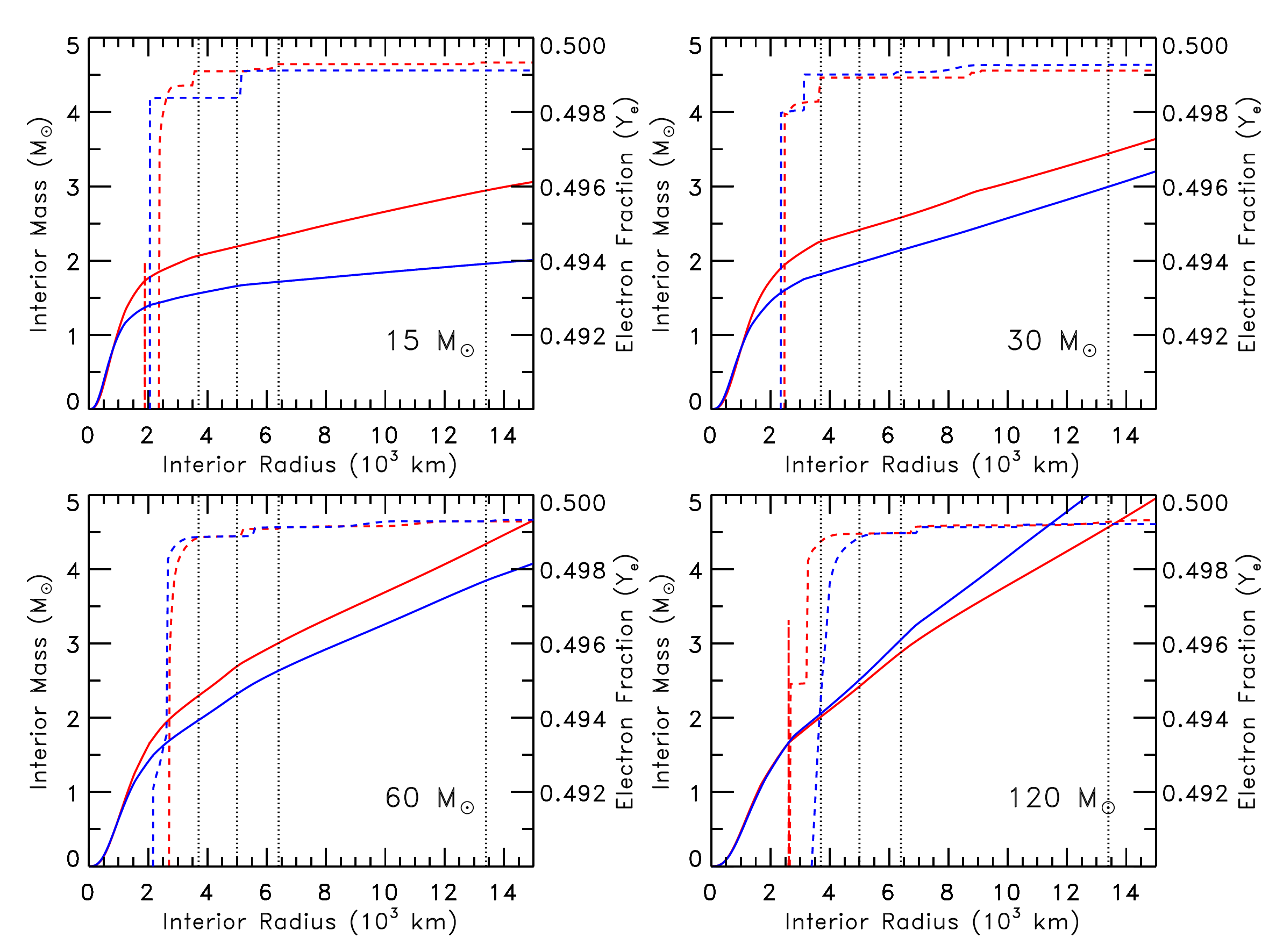}
\caption{Mass-Radius (M-R) relations of selected rotating (red lines) and non rotating (blue lines)
models in an advanced stage after core Si depletion characterized by similar M-R relations inside
$\sim 1~{\rm M_\odot}$ (see text).
Also shown in Figure is the electron fraction ($Y_s$) profile (dashed lines) which is reported on the
secondary y-axis.}
\label{mara4}
\end{figure}

In order to address the role of rotation on the advanced evolutionary phases we show in Figure \ref{omegahe0} the internal variation of both the angular velocity and the form factor ($f_P$) at two selected evolutionary stages after core He depletion, i.e., at core Si exhaustion (dotted lines) and at the presupernova stage (dashed lines). It is quite evident that, in spite of the spin up of the more internal zones due to the progressive contraction of the core, the structural deformations induced by rotation start to be sizable only after core Si exhaustion, i.e., the form factor (red lines) decreases substantially below 1 only beyond this phase. This means that all the advanced evolutionary phases (at least from core He exhaustion to core Si depletion) are not much affected by rotation and hence that the final structural differences between rotating and non rotating models depend, almost exclusively, on the differences in the CO core mass and in the central $\rm ^{12}C$ mass fraction at core He depletion. Let us briefly remind that the larger is the CO core - and/or the lower is the $\rm ^{12}C$ mass fraction at core He exhaustion - the faster is the advancing of the C shell and the more compact is the core of the star. According to these general rules, for the same initial mass, rotating models behave, in some sense, like more massive stars and hence end their life with more compact structures. Figures \ref{kip020R} and \ref{kip020N} show, as an example, a comparison between the Kippenhahn diagrams of a rotating and a non rotating 20 \msun model, respectively. While the non rotating model forms a convective core during core C burning, followed by three convective shell episodes in the further evolutionary phases (typical behavior of the less massive stars), the rotating model burns C radiatively in the core and presents only two C convective shell episodes in the further evolution (typical of a more massive star). The larger compactness of rotating models at the presupernova stage is shown in Figure \ref{mara4}, where the final Mass-Radius (M-R) relations of a subset of rotating models are compared to their respective non rotating counterparts.  It is important to remind, at this point, that the M-R relation plays a crucial role not only in the evolution of the innermost zones of the exploding mantle during the following explosion of the star (determining, e.g., the amount of falling back material), but also in determining the final explosive yields. The reason is that the volumes within which the various explosive burnings occur do not depend on the physical structure crossed by the shock wave and depend very modestly on the energy of the explosion. Therefore, the M-R relation determines essentially the amount of mass present in each one of these volumes and hence the amount of mass processed by the various explosive burnings \citep[see][for a detailed discussion of this topic]{cls00}. For this reason we show in Figure \ref{mara4} also the radii (vertical black dotted lines) identifying the zones undergoing the various explosive burnings for an explosion energy of 1 foe (1 foe = $10^{51}$ erg), for each selected presupernova model: from the left to the right the 4 vertical lines mark the distances at which the peak temperature of the shock wave drops to 5 GK (3700 km, limit of the complete explosive Si burning), 4 GK (5000 km, limit of the incomplete explosive Si burning), 3.3 GK (6400 km, limit of the explosive O burning) and 1.9 GK (13400 km, limit of the Ne/C explosive burnings). Another important quantity playing a crucial role in determining the isotope distribution of the explosive yields is the electron fraction ($Y_e$) profile at the presupernova stage. This quantity is mostly important for all the explosive burnings where a full- or a quasi- nuclear statistical equilibrium is achieved, i.e., complete and incomplete explosive Si burning and explosive O burning \citep{cls00}. While rotation leads to more compact final structures, it is worth noting that the $Y_e$ profile is not significantly affected by rotation, the largest (anyway small) difference occurring for the 15 \msun in the zone exposed to the explosive Si burning (see Figure \ref{mara4}). We will discuss the impact of these similarities and differences between rotating and non rotating models on the explosive yields in the next section.

\begin{figure}
\epsscale{.99}
\plotone{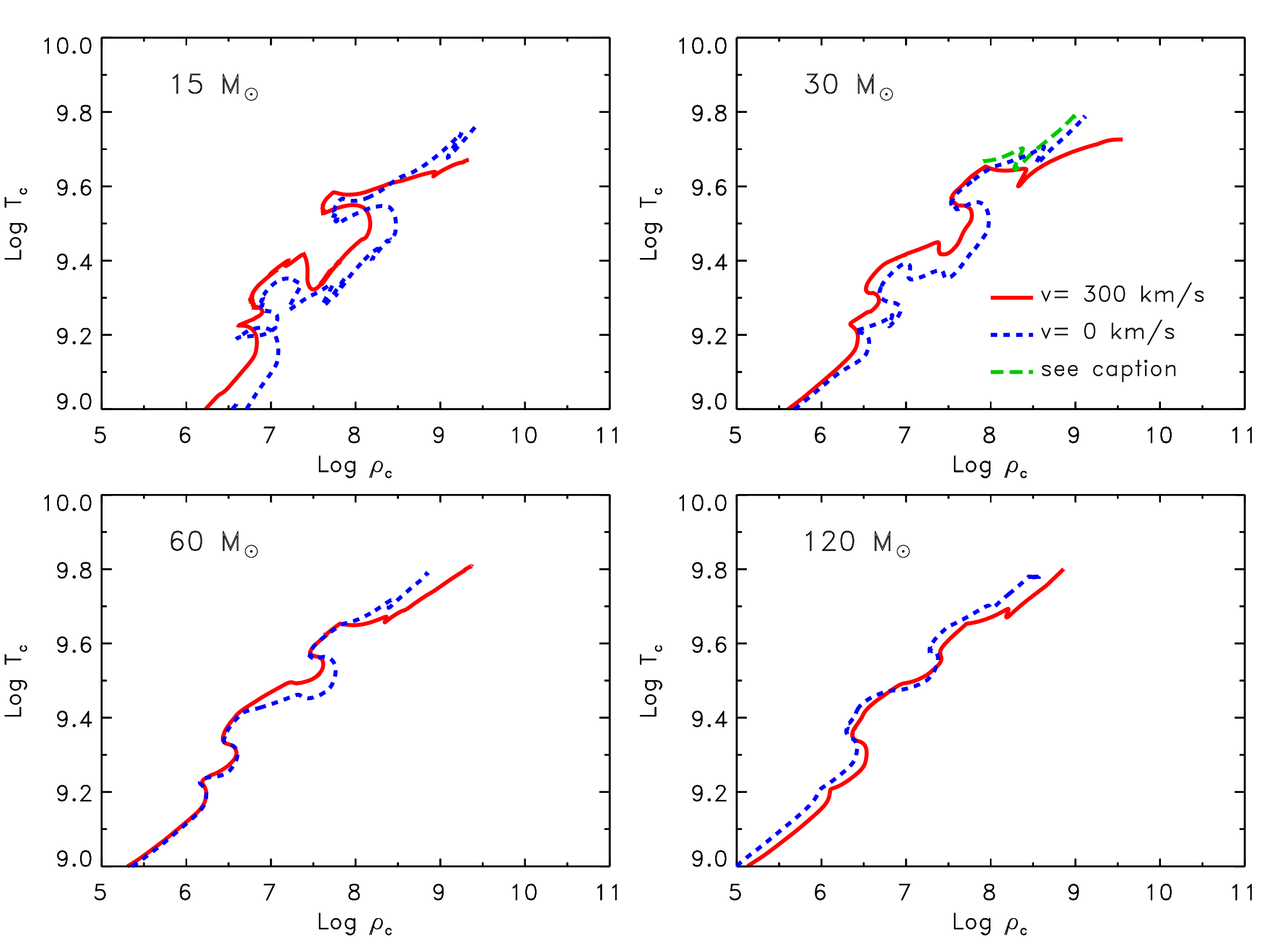}
\caption{Central temperature as a function of central density for four selected rotating (red solid lines) and non rotating (blue dotted lines) models.
The green long-dashed line in the upper right panel refers to a test model computed by imposing that both $f_P$ and $f_T$ could not decrease below 0.9}
\label{tcroc}
\end{figure}

The modest role played directly by rotation during the advanced burning phases becomes much more relevant beyond the core Si burning. Figure \ref{omegahe0} clearly shows that $f_P$ decreases significantly below 1 within the Fe core after the core Si depletion in both the 15 and 30 \msun (upper panels). By the way, let us remind again that our choice of adopting a constant initial surface equatorial velocity for all masses implies that the smaller the mass the larger the initial $\omega/\omega_{\rm crit}$ and hence the impact of rotation on the evolution of the stars. The large drop of $f_P$ implies a decrease of the effective gravity that in turn determines a slow down of the contraction. The cores of these rotating models therefore move towards a much more degenerate configuration compared to their non rotating counterparts. Figure \ref{tcroc} shows in fact that, while the evolutionary paths of the two sets of  models in the $\rm Log(T_c)-Log(\rho_c)$ plane remain quite similar up to the central Si depletion (the rotating models evolving at slightly larger entropies because of the systematically larger CO core masses), the rotating models definitely turn towards much higher densities and lower temperatures than their non rotating counterparts.

In order to really proof that this behavior is due to rotation, we recomputed the evolution of the 30 \msun from the core Si exhaustion onward, by imposing that  $f_P$ and $f_T$ could not decrease below 0.9. The resulting track, reported in Figure \ref{tcroc} as long dashed green line, shows that in this case the contraction of the core is not hindered anymore by rotation. This result is extremely interesting and leads to a possible speculation about the final evolution of fast rotating massive stars. If rotation were able to prevent the collapse of the core for a not negligible amount of time, the Fe core could grow much more than in the non rotating case before the onset of the collapse. Such an occurrence would hinder the passage of the shock wave because of the huge energy losses the shock wave undergoes inside the Fe core and, since the temporal delay would be controlled by the timescale of the outward leakage of angular momentum, a variety of different explosions could in principle occur. It would be very important to investigate in greater details the possible consequences of the stronger degeneracy we found to occur in fast rotating stellar models and we plan to address this issue in a forthcoming dedicated paper.

\begin{figure}
\epsscale{.99}
\plotone{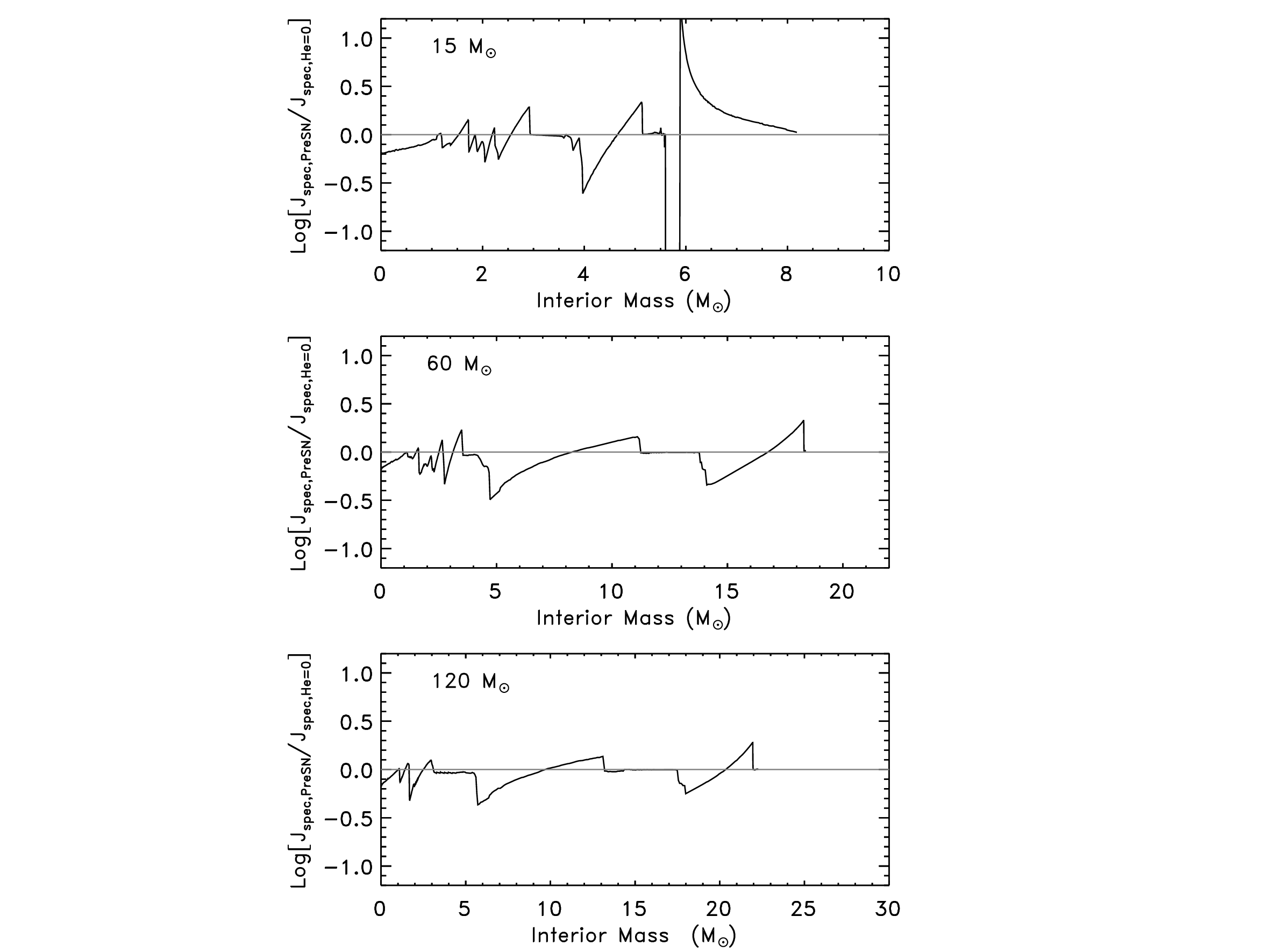}
\caption{Logarithm of the ratio between the final (at the presupernova stage) and initial (at core He exhaustion) specific angular momentum, as a function of the mass coordinate, for three selected models, namely, 15 (upper panel), 60 (middle panel) and 120 \msun 
(lower panel).}
\label{deltamangadv}
\end{figure}

Figure \ref{deltamangadv} shows the variation of the specific angular momentum between the end of core He burning and the presupernova stage. The {\it shark teeth} visible in the figure mark the convective regions where the rotational velocity is assumed to be flattened out (see Section 2) and hence the outward transport of the angular momentum maximized. In the radiative zone, viceversa, the specific angular momentum is, at a large extent, locally preserved because the speed up of the evolution due to the enormous neutrino energy losses largely inhibits both the secular shear and the meridional circulation. As a consequence, since no convective region crosses the border of the CO core, the total angular momentum locked in the CO core will remain basically constant up to the onset of the core collapse. 

The limiting masses that mark the passage from one kind of core collapse supernova to another are collected in Tables 5 and 6. For each progenitor mass (column 1), the amount of H and He present in the envelope at the time of the collapse are shown in columns 2 and 3 for rotating and non rotating models, respectively. The "kind" of star at the onset of the collapse is shown in column 4, i.e. whether the star is a RSG, YSG, or a WR, while the expected core collapse supernova type is shown in column 5. Note that, in order to classify the various core collapse supernovae we adopted the following H and He limiting masses \citep[see also][]{hachietal12}: (1) $\rm M(H)_{min,SNIIP}\simeq 0.3~M_\odot$ (the minimum H mass for a Type IIP SN); (2) $\rm M(H)_{min,SNIIb}\simeq 0.1~M_\odot$ (the minimum H mass for a Type IIb SN); (3) $\rm M(He)_{min,Ib}\simeq 0.1~M_\odot$ (the minimum He mass for a Type Ib SN). Inspection of the two mentioned tables show that in both cases, i.e. with and without rotation, the maximum mass for the explosion of a SNIIP is  $\rm M_{max,SNIIP}\simeq 17~M_\odot$. On the contrary, the minimum mass for the explosion of a SNIb is $\rm M_{min,SNIb}\simeq 30~M_\odot$ in the non rotating set while it lowers to $\rm M_{min,SNIb}\simeq 20~M_\odot$ if rotation with $v_{ini}(M_{\rm MS})=300~{\rm km/s}$ is taken into account. No progenitor star can produce a Type Ic SN lightcurve because of the rather high mass of He present in the envelope at the time of the explosion; note however that \cite{des11} question the upper limiting value of $0.1~M_\odot$ to have a Type Ic supernova explosion since they find that cases in which the He lines (in particular the He I 10830Ã line) are not excited even in presence of a quite large He mass fraction. Before closing this section let us note that the limiting masses recently obtained by the Geneva group \citep{georgy12} are in good agreement with the present findings.

\section{The explosive yields}

The explosion of the mantle of each model was followed by means of a hydrodynamic code developed by us that solves the fully compressible reactive hydrodynamic equations using the  Piecewise Parabolic Method (PPM) of \citet{cw84} in the lagrangean form. Each explosion was started by means of a kinetic bomb, i.e. by imparting instantaneously an initial velocity $v_{0}$ to a mass coordinate of $\sim 1~M_\odot$ , i.e. well within the iron core \citep{lc06}. 

Since these explosions are not obtained from \textit{first principles}, the initial velocity must be tuned in some way. In general, $v_{0}$ can be calibrated in order to obtain a given final kinetic energy of the ejecta or a given amount of mass ejected (i.e. a given amount of radioactive $\rm ^{56}Ni$). In the present paper we choose, for each model, the minimum initial velocity which provides the ejection of the whole mantle above the Fe core. This second case leaves the freedom to choose the  mass cut (i.e., the mass coordinate which separates the ejecta from the compact remnant) a posteriori, e.g. by requiring the ejection of a specific amount of $\rm ^{56}Ni$. Tables \ref{tabyields_rot} and \ref{tabyields_norot} show, for each exploded rotating and non rotating model, the Fe core mass (in solar masses), the kinetic energy of the ejecta resulting from the full ejection of the mass above the Fe core (in foes, i.e. in units of $10^{51}$ erg), the mass cut corresponding to the ejection of 0.1 \msun of $\rm ^{56}Ni$, the ejected mass of each nuclear species, in solar masses, corresponding to the ejection of 0.1 \msun of $\rm ^{56}Ni$. The evolution of the chemical composition during the explosion is followed up to a time of $t=2.5\cdot 10^{4}~\rm{s}$, so that all the unstable isotopes with a very short half-life have time to decay in this time interval. Let us eventually mention that, although the yields provided in Tables 7 and 8 have been obtained for a specific choice of the mass cut, the full set of cumulative isotopic yields as a function of the mass cut down to the Fe core mass are freely available at the Web site \url{http://www.iasf-roma.inaf.it/orfeo/public{\_}html}.

The comparison between the yields produced by rotating and non rotating models is not straightforward because, as already mentioned above, each single explosion must be tuned and we do not know if the same tuning should be applied to both sets of models. To be clearer, if we tune the explosion of the non rotating 15 \msun model in such a way that 0.1 \msun of $\rm ^{56}Ni$ is ejected, which kind of tuning should be used for the corresponding rotating model to make the comparison meaningful? And which is the role of the initial rotational velocity, i.e., how the tuning of the explosion should be changed with the initial rotational velocity? We do not have an answer and hence we will simply assume that both rotating and non rotating models eject the same amount of $\rm ^{56}Ni$. We are aware of the fact that probably this choice is not the best one, but this is the only approach we find viable at the moment. 

\begin{figure}
\epsscale{.99}
\plotone{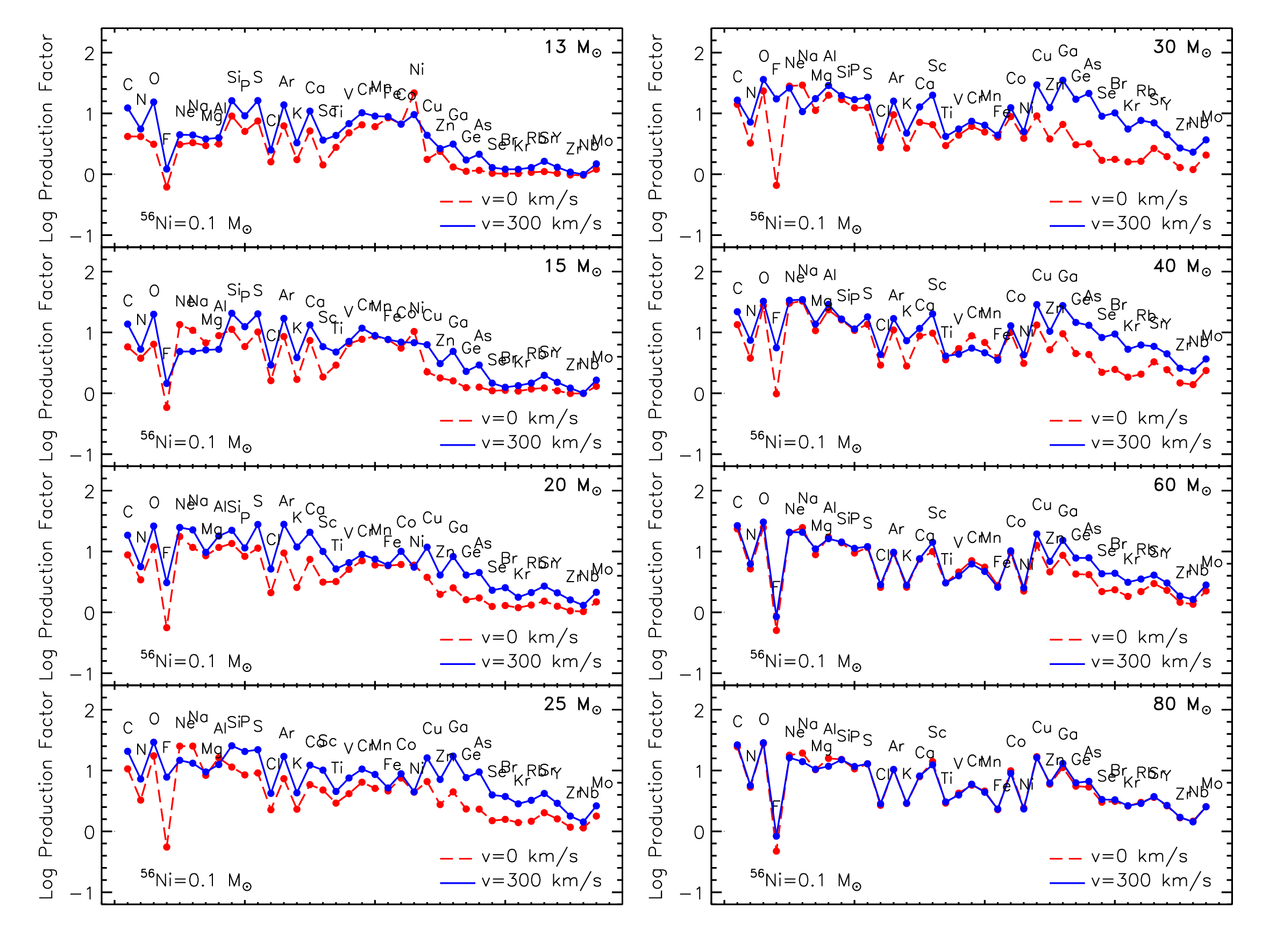}
\caption{Comparison between the element production factors of rotating (blue solid lines) and non
rotating (red dashed lines) models. The mass cut in all the modes has been fixed in order to obtain 0.1 $\rm M_\odot$ of $\rm ^{56}Ni$ in the ejecta.}
\label{pfni56}
\end{figure}


\begin{figure}
\epsscale{.99}
\plotone{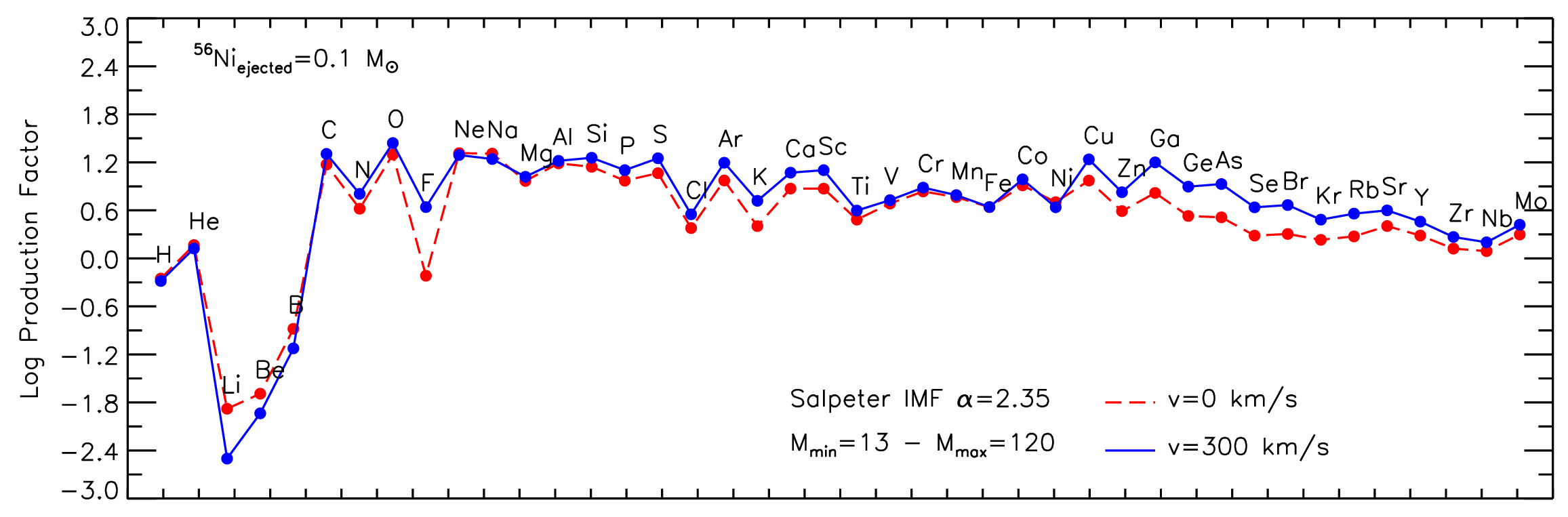}
\caption{Comparison between the element production factors, integrated over a Salpeter IMF (see text), of
rotating (blue solid lines) and non rotating (red dashed lines) models. The mass cut in all the modes has been fixed in order to obtain 0.1 $\rm M_\odot$ of $\rm ^{56}Ni$ in the ejecta.}
\label{pfimf}
\end{figure}

Figure \ref{pfni56} shows the comparison between the production factors for all the computed models, but the 120 \msun, obtained by assuming all the models to eject 0.1 $\rm M_\odot$ of $\rm ^{56}Ni$. The eight panels show that the overall differences are not large. However some interesting odds are present between the two sets of yields. In general, rotating models tend to overproduce, on average, all the elements, the ones with $M<40~{\rm M_\odot}$ preferentially the intermediate mass nuclei while the ones with $20\leq M/{\rm M_\odot} \leq 40$ the weak s-process component. Rotating stars with mass larger than 40 \msun, on the contrary, do not show any significant overproduction because the amount of angular momentum injected in these models is not enough to induce sizable effects on the final yields. A similar result is obtained by comparing the yields averaged over a Salpeter initial mass function $n(m)=k\cdot m^{-\alpha}$ (with $\alpha=2.35$). In this case, however, the differences are less pronounced than in the single mass comparisons (Figure \ref{pfimf}), although the depletion of Li and overproduction of both F and s-process elements obtained in the rotating models is still sizable.

As a final comment, let us point out that, as already discussed in the previous section, the rotating models are more compact than the non rotating ones, so that the binding energy of their mantle is systematically larger than the one of the corresponding non rotating counterparts. As a consequence, if we would assume for all models (rotating and not) the same final kinetic energy of the ejecta, e.g. 1 foe ($=10^{51}~\rm erg$), we would expect the rotating models to leave more massive remnants and to pollute less efficiently the interstellar medium. Such a possible scenario, however, could be reversed if at least part of the rotational energy could feed the outgoing shock wave: in this case one could have exactly the opposite result. 

\section{Summary and Conclusions}

In this paper we presented a first set of solar metallicity massive star models computed by taking into account the effects of rotation. The mechanical and thermal distortions induced by rotation have been included in our stellar evolutionary code (FRANEC) following the scheme proposed by \cite{kt70} and \cite{pin89} that can be considered a general approach \citep{hlw00,mm97}. We also included two rotation driven instabilities, namely the meridional circulation and the shear, following the schemes proposed either by \citet{mm03} (and references therein) and \citet{hlw00} (and references therein). Our results are similar to those obtained by other authors in the sense that the basic effects of rotation are similar, for similar initial conditions. They can be summarized as follows. 

Rotating stars (not too close to the break out velocity) spend most of their core H burning lifetime at lower luminosities and effective temperatures than their non rotating counterparts; the chemical mixing triggered by both the meridional circulation and the shear leads to larger H convective cores and also to a modification of the surface chemical composition. Since the average luminosity during core H burning is lower and the amount of fuel (H) is larger, the core H burning lifetime of the rotating models is longer than that of their non rotating counterparts. The amount of mass lost by the rotating stars is larger in most cases, with the noticeable exception of the mass interval between 60 and 80 \msun. The reason for a such a non monotonic effect of rotation on the mass loss is that the mass loss rate provided by \cite{val00} for OB stars is highly discontinuous in the proximity of two critical temperatures (bi-stability jumps), so that even minor changes in the path in the HR diagram may drastically change the amount of mass lost. 

Two important predictions are the temporal variations of the surface equatorial velocity and of the surface chemical composition (mainly N and He) because they have clear observational counterparts. A Large Program on the VLT with the FLAMES instrument (PI.: S.J. Smartt), focused on a survey of OB-type stars in the Galaxy and the Magellanic Clouds, started a few years ago. As part of this Large Program, a series of papers came out over the years presenting a large database of stars for which a quite detailed surface chemical abundances and projected rotational velocities ($v\sin(i)$) are now available \citep{untetal07,trundleetal07,untetal08,untetal09,dunstalletal11}. The comparison between theoretical predictions and these sample of stars in the "so called" Hunter diagram (i.e. $\epsilon(N)$ versus $v\sin(i)$) showed a significant discrepancy. We will not enter in this discussion here because most of the differences concern the Magellanic Cloud samples and here we presented solar metallicity stars. However, let us stress again that the present inclusion of rotation in a (any) evolutionary code is still very phenomenological and hence that the invocation of new exotic unknown phenomena could be untimely.

Rotation affects also the He burning phase, both directly and indirectly. On one side the rotating models, on the average, end the core H burning phase with smaller H rich mantles (but quite similar He core masses) than their corresponding non rotating models. This difference affects {\it per se} the core He burning because a) increases the time spent as BSG with respect to that spent as RSG, b) lowers the minimum mass that becomes a WR star, c) changes the initial final mass relation, d) changes the relative frequencies of the various core collapse supernova sub-types. On the other side, rotation triggers a partial mixing of matter between the He convective core and the surrounding He core (basically fresh $\rm ^{4}He$ is brought inside and fresh $\rm ^{12}C$ outside) and such an occurrence has the double consequence of lowering the amount of $\rm ^{12}C$ left by core He burning and of increasing the CO core mass. Moreover, the mixing of freshly synthesized $\rm ^{12}C$ from the He convective core to the more external He- and N-rich zones increases the C/N number ratio within the He core favoring the entrance in the WR-WNC stage when these zones are exposed to the surface. In the present set of rotating models ($v_{ini}(M_{\rm MS})=300~{\rm km/s}$), such a rotationally induced mixing is efficient enough that the number ratio C/N raises above 0.1 in the whole He core, therefore all the stars losing their H-rich envelope show up directly as WR-WNC skipping completely the WNE phase. During core He burning the outward transport of the angular momentum is efficient enough to reduce its internal initial (at core He ignition) value by about a fifty per cent. On the contrary, the structural distortions due to rotation during this phase are quite modest, i.e., both $f_P$ and $f_T$ never decrease significantly below 1. 

The evolution beyond core He exhaustion is characterized by the local conservation of the angular momentum because neither the meridional circulation nor the shear have time to operate any more. Angular momentum is redistributed only in the convective regions where the flattening of the angular velocity is assumed to occur on a dynamical timescale. From core He exhaustion to core Si depletion, both $f_P$ and $f_T$ never decrease significantly below 1, hence rotation does not play a considerable direct role during these stages. The only influence of rotation during these evolutionary phases is indirect, through the increase of the CO core mass and the decrease of the $\rm ^{12}C$ mass fraction at core He exhaustion. Both these occurrences have the effect of simulating the behavior of a more massive star, i.e., the mass-radius relation of any given rotating model at the presupernova stage is substantially steeper than the one of the corresponding non rotating counterpart. Since a steeper mass-radius relation means a larger gravitational binding energy, this result may have interesting consequences for the dynamics of the iron core collapse and fallback.

The influence of rotation on the final yields is difficult to assess because it is not clear how rotating and non rotating exploded models must be compared. If the comparison is made for the same final kinetic energy of the ejecta, we expect rotating models to leave more massive remnants and to pollute less than their non rotating counterparts (because of their larger compactness). Viceversa, if the comparison is made by assuming the ejection of the same amount of $\rm ^{56}Ni$, the chemical composition of the ejecta is quite similar with few differences worth being mentioned: (1) rotating models with mass in the range 20-40 \msun tend to overproduce by a factor of $\sim 2$ all the weak component s-process elements (Cu to Sr) because of their prolonged core He burning phase; (2) the same models also show a large overproduction of F, although its production factor remains substantially lower than that of the O; (3) the intermediate mass elements (S to Ca) are slightly enhanced together to O in the lower mass ($M<40~{\rm M_\odot}$) rotating models.

Let us note that the amount of angular momentum locked in the core would lead (if fully preserved) to a rotation velocity of the pulsar much larger (by one order of magnitude or so) than observed in the present available sample of millisecond pulsars \citep{kas94,mp96,mar98}. This is a well known problem \citep{hlw00,wh06}. The presently preferred possible candidate for an additional efficient transport of the angular momentum is the magnetic field: its role would be that of forcing a solid body rotation and hence maximize the efficiency of the outward transport of the angular momentum. Though the magnetic field could certainly play a pivotal role in this respect, our poor knowledge of the details of the collapse of a massive star and the following formation of the remnant leaves room for the possibility that a significant fraction of the angular momentum could be lost just during the collapse or the first phase of formation of the remnant from which a pulsar would emerge.

The strong influence rotation has on the amount of mass lost by a star during its life significantly affects the predicted limiting masses for the formation of the various WR subtypes as well as for the production of the various core collapse supernova types. In a simple (even if unrealistic) scenario in which all massive stars are born with the same surface equatorial velocity, one could easily redefine new limiting masses. For example, in a generation of massive stars with $v_{ini}(M_{\rm MS})=300~{\rm km/s}$ (our present models), the minimum mass that would show up as a Type Ib supernova would be of the order of 20 \msun. The corresponding value for a generation of non rotating stars would be of the order of 30 \msun. Also the various limiting masses for the WR subtypes would change as described in the previous sections. Actually we think that this approach is in principle misleading because the main characteristics of rotation is that it varies from one star to another and hence it must be seen as a new "space parameter". In other words rotation would simply spread all these limiting masses over a range of possible values. For example, a fast rotating 20 \msun may explode as a Type Ib supernova while another one, born with a smaller amount of angular momentum, may explode as a Type II-L Supernova. The same could hold for the initial mass-WR subtypes relation. If the distribution of the angular momentum among the different masses would be proved to follow a general rule, only in that case it would be possible to determine "average" new limiting masses. At present we feel safe to conclude that the limiting masses we obtained with and without rotation define the range of possible values for initial rotational velocities lower than 300 km/s.

Finally, we find that the maximum mass exploding as SNIIP is $\rm M_{max,SNIIP}\simeq 17~M_\odot$, independent of the inclusion of rotation. The substantial reduction of this quantity with respect to the value obtained up to now ($\sim 30~{\rm M_\odot}$) is due to the inclusion, in the present models, of the dust driven mass loss during the RSG stage \citep{vanloonetal05} which causes the star to lose a lot of mass during core He burning and to eventually enter the WR stage well before the explosion. This limiting value is in remarkable good agreement with the quite recent determinations of progenitor masses for SNIIP based on observed archival images provided by \citet{smartt09} (and references therein) which report a value of the order of $16\pm 0.5~{\rm M_\odot}$.

\begin{deluxetable}{lcccccccccccccccc}
\tabletypesize{\scriptsize}
\rotate
\setlength{\tabcolsep}{0.040in} 
\tablewidth{615.00000pt}
\tablecolumns{17}
\tablecaption{Main evolutionary properties of rotating models}
\tablehead{
 Phase &   Time   &  $\rm M_{CC}$  & Log($\rm T_{eff}$) & Log($\rm L/L_\odot$) &   M            & $\rm M_{He}$\tablenotemark{*} &    $\rm M_{CO}$    &   $\rm v_{equa}$   &  $\omega_{sup}$ &  $\omega / \omega_{crit}$ & $\rm J_{tot}$         &  $\rm H_{sup}$   &   $\rm He_{sup}$   &   $\rm N_{sup}$    &   N/C   &   N/O      \\
       &   (yr)   & ($\rm M_\odot$)&        (K)         &                      & ($\rm M_\odot$)& ($\rm M_\odot$)               &   ($\rm M_\odot$)  &  (km/s)            &  ($\rm s^{-1}$) &                           & $\rm 10^{53}$         &  mass            &  mass              &   mass             & number  & number    \\
       &          &                &                    &                      &                &                               &                    &                    &                 &                           & $\rm g~cm^{2}~s^{-1}$ &  fraction        &  fraction          &   fraction         & ratio   & ratio    }
\startdata
\cutinhead{13}
  H    & $1.87(+7)$ & $5.37$ &  $4.35$ &  $4.55$ &  $11.9$ &  $3.00$ &  $0.00$ &  $1.62(+2)$ &  $1.79(-05)$ &  $6.36(-1)$ &  $9.96(-2)$ &  $7.14(-1)$ &  $2.72(-1)$ &  $1.82(-3)$ &  $1.17(+0)$ &  $3.35(-1)$  \\ 
  He   & $1.72(+6)$ & $2.43$ &  $3.55$ &  $4.84$ &  $10.2$ &  $4.44$ &  $2.74$ &  $3.83(-1)$ &  $7.96(-10)$ &  $1.33(-2)$ &  $3.64(-2)$ &  $6.41(-1)$ &  $3.45(-1)$ &  $3.52(-3)$ &  $3.78(+0)$ &  $8.13(-1)$  \\ 
  C    & $3.13(+4)$ & $0.40$ &  $3.53$ &  $4.99$ &  $9.54$ &  $4.36$ &  $3.15$ &  $1.74(-1)$ &  $2.79(-10)$ &  $7.08(-3)$ &  $2.65(-2)$ &  $6.36(-1)$ &  $3.51(-1)$ &  $3.60(-3)$ &  $3.95(+0)$ &  $8.39(-1)$  \\ 
  Ne   & $2.27(+1)$ & $0.75$ &  $3.53$ &  $4.99$ &  $9.54$ &  $4.35$ &  $3.15$ &  $1.74(-1)$ &  $2.79(-10)$ &  $7.09(-3)$ &  $2.65(-2)$ &  $6.36(-1)$ &  $3.51(-1)$ &  $3.60(-3)$ &  $3.95(+0)$ &  $8.39(-1)$  \\ 
  O    & $1.92(+0)$ & $1.04$ &  $3.53$ &  $4.99$ &  $9.54$ &  $4.35$ &  $3.15$ &  $1.74(-1)$ &  $2.79(-10)$ &  $7.09(-3)$ &  $2.65(-2)$ &  $6.36(-1)$ &  $3.51(-1)$ &  $3.60(-3)$ &  $3.95(+0)$ &  $8.39(-1)$  \\ 
  Si   & $3.14(-1)$ & $1.38$ &  $3.53$ &  $4.99$ &  $9.54$ &  $4.35$ &  $2.88$ &  $1.74(-1)$ &  $2.79(-10)$ &  $7.08(-3)$ &  $2.65(-2)$ &  $6.36(-1)$ &  $3.51(-1)$ &  $3.60(-3)$ &  $3.95(+0)$ &  $8.39(-1)$  \\ 
 PSN   & $4.06(-3)$ &        &  $3.53$ &  $4.99$ &  $9.54$ &  $4.35$ &  $2.88$ &  $1.74(-1)$ &  $2.79(-10)$ &  $7.08(-3)$ &  $2.65(-2)$ &  $6.36(-1)$ &  $3.51(-1)$ &  $3.60(-3)$ &  $3.95(+0)$ &  $8.39(-1)$  \\
\enddata  
\tablenotetext{*}{$\rm M_{\rm He}=0$ means that the star has lost the whole H rich envelope and has become a bare He core}                                                                                                                                                                                                                                                                                                                                                            
\end{deluxetable}

\begin{deluxetable}{lcccccccccccc}
\tabletypesize{\scriptsize}
\rotate
\setlength{\tabcolsep}{0.04in} 
\tablecolumns{13}
\tablewidth{465.00000pt}
\tablecaption{Main evolutionary properties of non rotating models}
\tablehead{
 Phase &   Time   &  $\rm M_{CC}$  & Log($\rm T_{eff}$) & Log($\rm L/L_\odot$) &   M            &    $\rm M_{He}$\tablenotemark{*} &    $\rm M_{CO}$         &    $\rm H_{sup}$   &   $\rm He_{sup}$   &   $\rm N_{sup}$    &   N/C     &   N/O   \\
       &   (yr)   & ($\rm M_\odot$)&        (K)         &                      & ($\rm M_\odot$)& ($\rm M_\odot$)                  &   ($\rm M_\odot$)  &    mass            &  mass              &   mass             & number  & number    \\
       &          &                &                    &                      &                &                                  &                    &    fraction        &  fraction          &   fraction         & ratio   & ratio     }  
\startdata
\cutinhead{13}
  H   & $1.61(+7)$ & $5.47$ &  $4.39$ &  $4.53$ &  $12.9$ &  $2.90$ &  $0.00$ &   $7.21(-1)$ &  $2.65(-1)$ &  $6.95(-04)$ &  $2.96(-01)$ &  $1.21(-01)$ \\
  He  & $1.03(+6)$ & $1.47$ &  $3.57$ &  $4.66$ &  $12.2$ &  $4.06$ &  $1.49$ &   $7.02(-1)$ &  $2.84(-1)$ &  $2.37(-03)$ &  $1.69(+00)$ &  $4.70(-01)$ \\
  C   & $3.48(+4)$ & $0.70$ &  $3.56$ &  $4.79$ &  $12.0$ &  $4.07$ &  $1.91$ &   $7.02(-1)$ &  $2.84(-1)$ &  $2.37(-03)$ &  $1.69(+00)$ &  $4.70(-01)$ \\
  Ne  & $6.51(+1)$ & $0.66$ &  $3.56$ &  $4.79$ &  $12.0$ &  $4.07$ &  $1.88$ &   $7.02(-1)$ &  $2.84(-1)$ &  $2.37(-03)$ &  $1.69(+00)$ &  $4.70(-01)$ \\
  O   & $5.52(+0)$ & $0.90$ &  $3.56$ &  $4.79$ &  $12.0$ &  $4.07$ &  $1.88$ &   $7.02(-1)$ &  $2.84(-1)$ &  $2.37(-03)$ &  $1.69(+00)$ &  $4.70(-01)$ \\
  Si  & $1.10(+0)$ & $0.97$ &  $3.56$ &  $4.79$ &  $12.0$ &  $4.07$ &  $1.88$ &   $7.02(-1)$ &  $2.84(-1)$ &  $2.37(-03)$ &  $1.69(+00)$ &  $4.70(-01)$ \\
 PSN  & $2.47(-3)$ &        &  $3.56$ &  $4.79$ &  $12.0$ &  $4.07$ &  $1.81$ &   $7.02(-1)$ &  $2.84(-1)$ &  $2.37(-03)$ &  $1.69(+00)$ &  $4.70(-01)$ \\ 
\cutinhead{15}                                                                                                                                                                                                
  H   & $1.31(+7)$ & $6.64$ &  $4.42$ &  $4.71$ &  $14.9$ &  $3.64$ &  $0.00$ &   $7.21(-1)$ &  $2.65(-1)$ &  $6.95(-04)$ &  $2.96(-01)$ &  $1.21(-01)$ \\
\enddata                                                                                                                                                                                                                                                                                                                              
\tablenotetext{*}{$\rm M_{\rm He}=0$ means that the star has lost the whole H rich envelope and has become a bare He core}
\end{deluxetable}

\begin{deluxetable}{ccccccccccc}
\tabletypesize{\scriptsize}
\rotate
\tablecolumns{11}
\tablewidth{0pc}
\tablecaption{Wolf-Rayet lifetimes of rotating models}
\tablehead{
 Initial Mass   &   $t_{\rm O}$   & $t_{\rm WR}$   &   $t_{\rm WNL}$  &  $\rm H_c/He_c÷(WNL)$   &  $t_{\rm WNE}$  & $\rm H_c/He_c÷(WNE)$ & $t_{\rm WNC}$ & $\rm H_c/He_c÷(WNC)$ & $t_{\rm WCO}$ & $\rm H_c/He_c÷(WCO)$  \\
($\rm M_\odot$) &     (yr)        & (yr)            &   (yr)          &  mass fraction          &  (yr)           & mass fraction        &  (yr)           & mass fraction        &  (yr)       & mass fraction        }
\startdata
  13  &                       &                      &                      &          &                         &          &                      &          &                     &          \\
  15  &   $2.18\cdot 10^{4}$  &                      &                      &          &                         &          &                      &          &                     &          \\
  20  &   $4.79\cdot 10^{6}$  &  $3.44\cdot 10^{5}$  &  $2.93\cdot 10^{5}$  & He(0.29) &                         &          &  $5.10\cdot 10^{4}$  & He(0.02) &                     &          \\
  25  &   $5.70\cdot 10^{6}$  &  $1.54\cdot 10^{5}$  &  $1.17\cdot 10^{5}$  & He(0.12) &                         &          &  $3.74\cdot 10^{4}$  & He(0.01) &                     &          \\
  30  &   $5.30\cdot 10^{6}$  &  $2.33\cdot 10^{5}$  &  $2.03\cdot 10^{5}$  & He(0.25) &                         &          &  $3.00\cdot 10^{4}$  & He(0.01) &                     &          \\
  40  &   $4.55\cdot 10^{6}$  &  $4.16\cdot 10^{5}$  &  $2.75\cdot 10^{5}$  & He(0.74) &                         &          &  $4.13\cdot 10^{4}$  & He(0.17) &  $9.98\cdot 10^{4}$ & He(0.10) \\
  60  &   $3.73\cdot 10^{6}$  &  $4.01\cdot 10^{5}$  &  $1.54\cdot 10^{5}$  & He(0.95) &                         &          &  $1.20\cdot 10^{4}$  & He(0.48) &  $2.35\cdot 10^{5}$ & He(0.45) \\
  80  &   $3.39\cdot 10^{6}$  &  $3.60\cdot 10^{5}$  &  $7.29\cdot 10^{4}$  & He(0.99) &                         &          &  $1.24\cdot 10^{4}$  & He(0.70) &  $2.75\cdot 10^{5}$ & He(0.65) \\
\enddata
\end{deluxetable}

\begin{deluxetable}{ccccccccccc}
\tabletypesize{\scriptsize}
\rotate
\tablecolumns{11}
\tablewidth{0pc}
\tablecaption{Wolf-Rayet lifetimes of non rotating models}
\tablehead{
 Initial Mass &   $t_{\rm O}$   & $t_{\rm WR}$   &   $t_{\rm WNL}$  &  $\rm H_c/He_c÷(WNL)$   &  $t_{\rm WNE}$  & $\rm H_c/He_c÷(WNE)$ & $t_{\rm WNC}$ & $\rm H_c/He_c÷(WNC)$ & $t_{\rm WCO}$ & $\rm H_c/He_c÷(WCO)$  \\
 ($\rm M_\odot$) &     (yr)        & (yr)            &   (yr)          &  mass fraction          &  (yr)           & mass fraction        &  (yr)           & mass fraction        &  (yr)       & mass fraction        }
\startdata
  13  &                        &                      &                      &          &                      &          &                      &          &                     &          \\
  15  &   $1.07\cdot 10^{5}$   &                      &                      &          &                      &          &                      &          &                     &          \\
  20  &   $6.28\cdot 10^{6}$   &                      &                      &          &                      &          &                      &          &                     &          \\
  25  &   $5.80\cdot 10^{6}$   &  $1.07\cdot 10^{4}$  &  $1.07\cdot 10^{4}$  & Adv.     &                      &          &                      &          &                     &          \\
  30  &   $5.14\cdot 10^{6}$   &  $1.10\cdot 10^{5}$  &  $1.10\cdot 10^{5}$  & He(0.12) &                      &          &                      &          &                     &          \\
  40  &   $4.18\cdot 10^{6}$   &  $1.64\cdot 10^{5}$  &  $1.35\cdot 10^{5}$  & He(0.30) &  $2.89\cdot 10^{4}$  & He(0.02) &                      &          &                     &          \\
  60  &   $3.20\cdot 10^{6}$   &  $3.59\cdot 10^{5}$  &  $7.79\cdot 10^{4}$  & He(0.97) &  $1.58\cdot 10^{5}$  & He(0.67) &  $1.84\cdot 10^{4}$  & He(0.20) &  $1.05\cdot 10^{5}$ & He(0.16) \\
  80  &   $2.62\cdot 10^{6}$   &  $3.67\cdot 10^{5}$  &  $9.54\cdot 10^{4}$  & H(0.01)  &  $9.94\cdot 10^{4}$  & He(0.74) &  $1.64\cdot 10^{4}$  & He(0.38) &  $1.56\cdot 10^{5}$ & He(0.32) \\
\enddata
\end{deluxetable}

\begin{deluxetable}{cccccc}
\tablecolumns{5}
\tablewidth{0pc}
\tablecaption{Selected properties of rotating models at the presupernova stage}
\tablehead{
 Initial Mass    &   H             & He              &   Configuration  &  SN-Type  \\
 ($\rm M_\odot$) & ($\rm M_\odot$) & ($\rm M_\odot$) &                  &           }
\startdata
  13  & 3.18  & 2.77  & RSG  & SNIIP \\
  15  & 1.80  & 2.18  & RSG  & SNIIP \\
  20  & 0.00  & 0.74  & WNC  & SNIb  \\
  25  & 0.00  & 0.76  & WNC  & SNIb  \\
  30  & 0.00  & 0.78  & WNC  & SNIb  \\
  40  & 0.00  & 0.57  & WCO  & SNIb  \\
  60  & 0.00  & 0.45  & WCO  & SNIb  \\
  80  & 0.00  & 0.45  & WCO  & SNIb  \\
 120  & 0.00  & 0.63  & WCO  & SNIb 
\enddata
\end{deluxetable}

\begin{deluxetable}{cccccc}
\tablecolumns{5}
\tablewidth{0pc}
\tablecaption{Selected properties of non rotating models at the presupernova stage}
\tablehead{
 Initial Mass &   H   & He   &   Configuration  &  SN-Type   \\  
  ($\rm M_\odot$) & ($\rm M_\odot$) & ($\rm M_\odot$) &                  &           }
\startdata
  13  & 5.46  & 4.49  & RSG  & SNIIP \\
  15  & 5.76  & 4.91  & RSG  & SNIIP \\
  20  & 0.26  & 3.83  & RSG  & SNIIb  \\
  25  & 0.16  & 4.45  & YSG  & SNIIb  \\
  30  & 0.03  & 4.33  & YSG  & SNIb  \\
  40  & 0.00  & 4.20  & WNE  & SNIb  \\
  60  & 0.00  & 0.92  & WCO  & SNIb  \\
  80  & 0.00  & 0.80  & WCO  & SNIb  \\
 120  & 0.00  & 0.54  & WCO  & SNIb 
\enddata
\end{deluxetable}

\begin{deluxetable}{lccccccccc}
\tabletypesize{\scriptsize}
\setlength{\tabcolsep}{0.05in}
\tablecolumns{10}
\tablewidth{0pc}
\tablecaption{Explosion properties and ejected masses (in $\rm M_\odot$ and $\rm 2.5\cdot 10^4~s$ after the explosion) of the rotating models \label{tabyields_rot}}
\tablehead{
\colhead{}     & \colhead{$\rm 	13$}& \colhead{$\rm 15$} & \colhead{$\rm 20$} & \colhead{$\rm 25$} & \colhead{$\rm 30$} & \colhead{$\rm 40$} & \colhead{$\rm 60$} & \colhead{$\rm 80$} & \colhead{$\rm 120$}      }
\startdata
$\rm M_{Fe} $  &     1.73           &    1.85            &     1.83           &     2.03           &    2.14            &    2.00            &     2.27           &     2.17           &     1.80                 \\
$\rm E_{kin}$  &     1.00           &    1.00            &     1.17           &     1.91           &    2.05            &    2.21            &     7.39           &     6.55           &     2.91                 \\
$\rm M_{cut}$  &     1.84           &    2.02            &     1.92           &     2.42           &    2.37            &    2.33            &     2.87           &     3.11           &     2.68                 \\
$\rm ^{ 1}H $  &  $\rm  5.48(+00)$  &  $\rm  6.06(+00)$  &  $\rm  7.59(+00)$  &  $\rm  9.11(+00)$  &  $\rm  9.88(+00)$  &  $\rm  1.32(+01)$  &  $\rm  1.89(+01)$  &  $\rm  2.43(+01)$  &  $\rm  3.70(+01)$ \\          
$\rm ^{ 2}H $  &  $\rm  1.40(-08)$  &  $\rm  1.66(-16)$  &  $\rm  1.92(-16)$  &  $\rm  2.17(-16)$  &  $\rm  2.28(-16)$  &  $\rm  2.65(-16)$  &  $\rm  3.02(-16)$  &  $\rm  3.67(-16)$  &  $\rm  5.00(-16)$ \\          
$\rm ^{ 3}H $  &  $\rm  1.87(-31)$  &  $\rm  2.35(-31)$  &  $\rm  4.33(-31)$  &  $\rm  3.03(-30)$  &  $\rm  1.36(-30)$  &  $\rm  2.40(-30)$  &  $\rm  7.03(-30)$  &  $\rm  1.25(-29)$  &  $\rm  1.39(-29)$ \\          
$\rm ^{ 3}He$  &  $\rm  2.38(-04)$  &  $\rm  2.57(-04)$  &  $\rm  3.27(-04)$  &  $\rm  4.01(-04)$  &  $\rm  4.28(-04)$  &  $\rm  5.87(-04)$  &  $\rm  8.99(-04)$  &  $\rm  1.12(-03)$  &  $\rm  1.54(-03)$ \\          
$\rm ^{ 4}He$  &  $\rm  3.87(+00)$  &  $\rm  4.38(+00)$  &  $\rm  5.52(+00)$  &  $\rm  7.08(+00)$  &  $\rm  8.70(+00)$  &  $\rm  1.22(+01)$  &  $\rm  2.11(+01)$  &  $\rm  3.09(+01)$  &  $\rm  5.81(+01)$ \\
\enddata         
\end{deluxetable}

\begin{deluxetable}{lccccccccc} 
\tabletypesize{\scriptsize}
\setlength{\tabcolsep}{0.05in}
\tablecolumns{10}
\tablewidth{0pc}
\tablecaption{Explosion properties and ejected masses (in $\rm M_\odot$ and $\rm 2.5\cdot 10^4~s$ after the explosion) of the non rotating models \label{tabyields_norot}}
\tablehead{
\colhead{} & \colhead{$\rm 	13$}   & \colhead{$\rm 15$}  & \colhead{$\rm 20$}    & \colhead{$\rm 25$} &
\colhead{$\rm 30$} & \colhead{$\rm 40$}   & \colhead{$\rm 60$}  & \colhead{$\rm 80$}  & \colhead{$\rm 120$} }
\startdata
$\rm M_{Fe} $  &     1.14                    &    1.40                     &     1.40                    &     1.48                    &    1.56                     &    1.58                     &     1.48                    &     1.75                    &     1.92                           \\
$\rm E_{kin}$  &     1.00                    &    1.00                     &     1.00                    &     1.15                    &    1.63                     &    1.77                     &     4.22                    &     4.32                    &     5.48                           \\
$\rm M_{cut}$  &     1.38                    &    1.42                     &     1.57                    &     1.65                    &    1.79                     &    2.08                     &     2.21                    &     2.43                    &     2.81                           \\
$\rm ^{ 1}H $  &  $\rm  6.15(+00)$  &  $\rm  6.87(+00)$  &  $\rm  8.46(+00)$  &  $\rm  1.00(+01)$  &  $\rm  1.13(+01)$  &  $\rm  1.43(+01)$  &  $\rm  2.04(+01)$  &  $\rm  2.51(+01)$  &  $\rm  3.58(+01)$         \\
$\rm ^{ 2}H $  &  $\rm  3.59(-06)$  &  $\rm  3.67(-06)$  &  $\rm  5.91(-06)$  &  $\rm  7.40(-06)$  &  $\rm  8.90(-06)$  &  $\rm  3.14(-16)$  &  $\rm  3.98(-16)$  &  $\rm  4.36(-16)$  &  $\rm  1.76(-04)$         \\
$\rm ^{ 3}H $  &  $\rm  1.51(-17)$  &  $\rm  7.39(-17)$  &  $\rm  4.35(-12)$  &  $\rm  5.04(-12)$  &  $\rm  5.70(-12)$  &  $\rm  1.30(-30)$  &  $\rm  4.25(-30)$  &  $\rm  9.38(-30)$  &  $\rm  1.03(-29)$         \\
$\rm ^{ 3}He$  &  $\rm  3.54(-04)$  &  $\rm  3.82(-04)$  &  $\rm  4.45(-04)$  &  $\rm  5.24(-04)$  &  $\rm  5.71(-04)$  &  $\rm  7.22(-04)$  &  $\rm  1.12(-03)$  &  $\rm  1.35(-03)$  &  $\rm  1.53(-03)$         \\
$\rm ^{ 4}He$  &  $\rm  4.77(+00)$  &  $\rm  5.35(+00)$  &  $\rm  7.31(+00)$  &  $\rm  8.97(+00)$  &  $\rm  1.00(+01)$  &  $\rm  1.35(+01)$  &  $\rm  2.26(+01)$  &  $\rm  3.12(+01)$  &  $\rm  5.04(+01)$         \\
\enddata                                  
\end{deluxetable}


\end{document}